\documentclass[a4paper,fleqn,usenatbib]{mnras}

\usepackage{newtxtext,newtxmath}

\usepackage[T1]{fontenc}
\usepackage{ae,aecompl}

\usepackage{graphicx}	
\usepackage{amsmath}	

\newcommand{\ngc}[1]{\ifnum #1=6093 M80\else NGC~{#1}\fi}
\newcommand{\changed}[1]{{#1}}
\newcommand{\changedtwo}[1]{{#1}}
\newcommand{\msol}{\text{M}_\odot}
\newcommand{\lsol}{\text{L}_\odot}
\newcommand{\kms}{$\text{km}\,\text{s}^{-1}$}
\newcommand{\hst}{\textit{HST}}

\title[Central kinematics of M80]{Central kinematics of the Galactic globular cluster M80}

\author[F. Göttgens]{
Fabian Göttgens,$^{1}$\thanks{E-mail: fabian.goettgens@uni-goettingen.de}
Sebastian Kamann,$^{2}$
Holger Baumgardt,$^{3}$
Stefan Dreizler,$^{1}$
\newauthor{
Benjamin Giesers,$^{1}$
Tim-Oliver Husser,$^{1}$ 
Mark den Brok,$^{4}$
Romain Fétick,$^{5,\,6}$
}
\newauthor{
Davor Krajnović,$^{4}$
and Peter M.\ Weilbacher$^{4}$
}
\\
$^{1}$Institut für Astrophysik, Georg-August-Universität Göttingen, Friedrich-Hund-Platz 1, 37077 Göttingen, Germany\\
$^{2}$Astrophysics Research Institute, Liverpool John Moores University, 146 Brownlow Hill, Liverpool L3 5RF, UK\\
$^{3}$School of Mathematics and Physics, The University of Queensland, St. Lucia, QLD 4072, Australia\\
$^{4}$Leibniz-Institut für Astrophysik Potsdam (AIP), An der Sternwarte 16, 14482 Potsdam, Germany \\
$^{5}$ DOTA, ONERA, Université Paris Saclay, F-92322 Châtillon, France \\
$^{6}$ Aix-Marseille Université, CNRS, CNES, LAM, Marseille, France
}

\date{Accepted XXX. Received YYY; in original form ZZZ}

\pubyear{2021}

\begin{document}
\label{firstpage}
\pagerange{\pageref{firstpage}--\pageref{lastpage}}
\maketitle

\begin{abstract}

We use spectra observed with the integral-field spectrograph MUSE to reveal the central kinematics of the Galactic globular cluster Messier 80 (M80, NGC~6093). 
Using observations obtained with the recently commissioned narrow-field mode of MUSE, we are able to analyse 932 stars in the central 7.5~arcsec by 7.5~arcsec of the cluster for which no useful spectra previously existed.
Mean radial velocities of individual stars derived from the spectra are compared to predictions from axisymmetric Jeans models, resulting in radial profiles of the velocity dispersion, the rotation amplitude, and the mass-to-light ratio. The new data allow us to search for an intermediate-mass black hole (IMBH) in the centre of the cluster. 
Our Jeans model finds two similarly probable solutions around different dynamical cluster centres. The first solution has a centre close to the photometric estimates available in the literature and does not need an IMBH to fit the observed kinematics. 
The second solution contains a location of the cluster centre that is offset by about 2.4~arcsec from the first one and it needs an IMBH mass of $4600^{+1700}_{-1400}~\msol{}.$
$N$-body models support the existence of an IMBH \changedtwo{in this cluster} with a mass of up to $6000~\msol$ in this cluster, \changedtwo{although models without an IMBH provide a better fit to the observed surface brightness profile. They further} indicate that the cluster has lost nearly all stellar-mass black holes.
We further discuss the detection of two potential high-velocity stars with radial velocities of 80 to 90~\kms{} relative to the cluster mean.

\end{abstract}



\section{Introduction}
The cores of globular clusters (GCs) are among the regions with the highest density of stars.
With up to $10^5$ stars per cubic parsec, they contain a multitude of stellar exotica including millisecond pulsars, rejuvenated and heavy stars in form of blue stragglers, cataclysmic variables, accreting and non-accreting stellar-mass black holes as remnants of high-mass stars, and potentially intermediate-mass black holes (IMBH).
IMBHs are a hypothetical class of black holes which would fill the gap between stellar-mass black holes with up to a few tens of solar masses and super-massive black holes in the centres of galaxies with masses ranging from millions to billions of solar masses \citep[][]{greene_intermediate-mass_2020}. 
Numerical simulations of globular clusters predict that they can contain a single IMBH with a mass of several thousand solar masses~\citep{arca_sedda_mocca-survey_2019}, formed by merging stellar-mass black holes with massive stars and binaries or by merging massive stars \citep[e.g.][]{portegies_zwart_formation_2004,giersz_mocca_2015,rizzuto_intermediate_2020}.
An IMBH could be detected through its influence on stellar kinematics as it will increase the velocity dispersion of stars in the centre of a globular cluster.
If there is enough gas in the core of a GC, it could be accreted by the IMBH which could cause detectable radio or X-ray emission if the overall accretion efficiency is not too low \citep{tremou_maveric_2018}.
The deep gravitational potential of an IMBH should cause a change in the period of millisecond pulsars, which could be detectable if all other accelerations are accurately modelled \citep{kiziltan_intermediate-mass_2017,henault-brunet_black_2020,abbate_evidence_2019}.
So far, observations have not resulted in a convincing detection of an IMBH in a globular cluster but yielded upper mass limits \citep[see Table 3 in][]{greene_intermediate-mass_2020}. 
The recent gravitational-wave signal GW190521 points to a heavy BH remnant with a mass of approx. $140~\msol{}$ \citep{ligo_scientific_collaboration_and_virgo_collaboration_gw190521_2020,abbott_properties_2020}, which is at the low-mass end of IMBHs. 
It is unclear in which environment this merger occurred, but star clusters provide favourable conditions \changed{but star clusters provide favourable conditions for the merging of black holes that are more massive than what is predicted by single-star evolution} \citep{abbott_properties_2020}. 

A challenge in the detection of IMBHs via stellar kinematics lies in the unknown amount and mass of stellar remnants residing near the cluster centres \citep[e.g. ][]{gieles_mass_2018,zocchi_effect_2019,mann_multimass_2019,baumgardt_no_2019,henault-brunet_black_2020}.
However, detailed studies of the central cluster kinematics can help to discriminate between the presence of an IMBH and an overdensity of stellar remnants.
In case of $\omega$~Cen (\ngc{5139}), \citet{baumgardt_no_2019} found using N-body models that the proposed $\approx 45{,}000~\msol{}$ IMBH \citep{noyola_very_2010, jalali_dynamical_2012, baumgardt_n-body_2017} would produce about 20 stars with a radial velocity above the maximum stellar velocity measured in the centre of this cluster, while models with a number of stellar-mass black holes instead of a central IMBH do not contain these high-velocity stars.
IMBHs are statistically expected to have companion stars inside their sphere of influence, which would extend over several tenths of arcseconds to a few arcseconds for most clusters.
Although the innermost companion of an IMBH in a 10-Gyr-old GC will typically be a neutron star, a massive ($> 1~\msol{}$) white dwarf, or a stellar-mass black hole \citep{macleod_close_2016}, which are impossible to observe visually, other companions might be observable. 
Interactions between the IMBH and stars, e.g. a fly-by, could accelerate stars and cause a proper motion or radial velocity much higher than those of stars that did not interact with the IMBH.

M80 (NGC~6093) is an old Milky Way globular cluster with an age of $13.5 \pm 1.0$~Gyr \citep{dotter_acs_2009}. It is located in the direction of the Galactic centre at a \changed{heliocentric} distance of $8.86 \pm 0.55$~kpc \citep{baumgardt_catalogue_2018}. The cluster core radius is $r_c \approx 0.36\,\text{pc} = 8$~arcsec \citep{harris_catalog_1996,baumgardt_catalogue_2018}.
It belongs to the group of dynamically old globular clusters \citep{ferraro_dynamical_2012}.
While scaling relations derived from GC simulations predict an IMBH with a mass of $(3.63 \pm 0.95) \times 10^3 \text{M}_\odot$ in M80 \citep{arca_sedda_mocca-survey_2019}, an integrated-light study did not find evidence for an IMBH \citep{lutzgendorf_limits_2013}. 
\cite{kamann_peculiar_2020} studied the kinematics of M80 using radial velocities derived from MUSE data from 2015--2017 and the axisymmetric Jeans model from \citet{watkins_discrete_2013}. This study found that stars belonging to different chemical populations \citep[as discovered by][]{dalessandro_peculiar_2018} rotate differently.

In this paper, we build on the study of \cite{kamann_peculiar_2020} and use new data from the centremost stars in the cluster obtained with the recently commissioned narrow-field mode of MUSE. We combine this with new MUSE wide-field mode data and updated radial velocity estimates (see Sections~\ref{sect:observations} and \ref{sect:extraction}) to 
re-analyse the global kinematics of M80 with a Jeans model and calculate rotation profiles, mass-to-light ratio profiles and infer the mass of a hypothetical IMBH (Section~\ref{sect:model_upper}). 
Section~\ref{sect:nbody} presents results derived from $N$-body models and the stars with high radial velocities are described in Section~\ref{sct:highvrad_stars}. We discuss our results in Section~\ref{sect:discussion} and conclude in Section~\ref{sect:conclusion}.

\section{MUSE Observations}
\begin{figure*}
	\includegraphics[width=2\columnwidth]{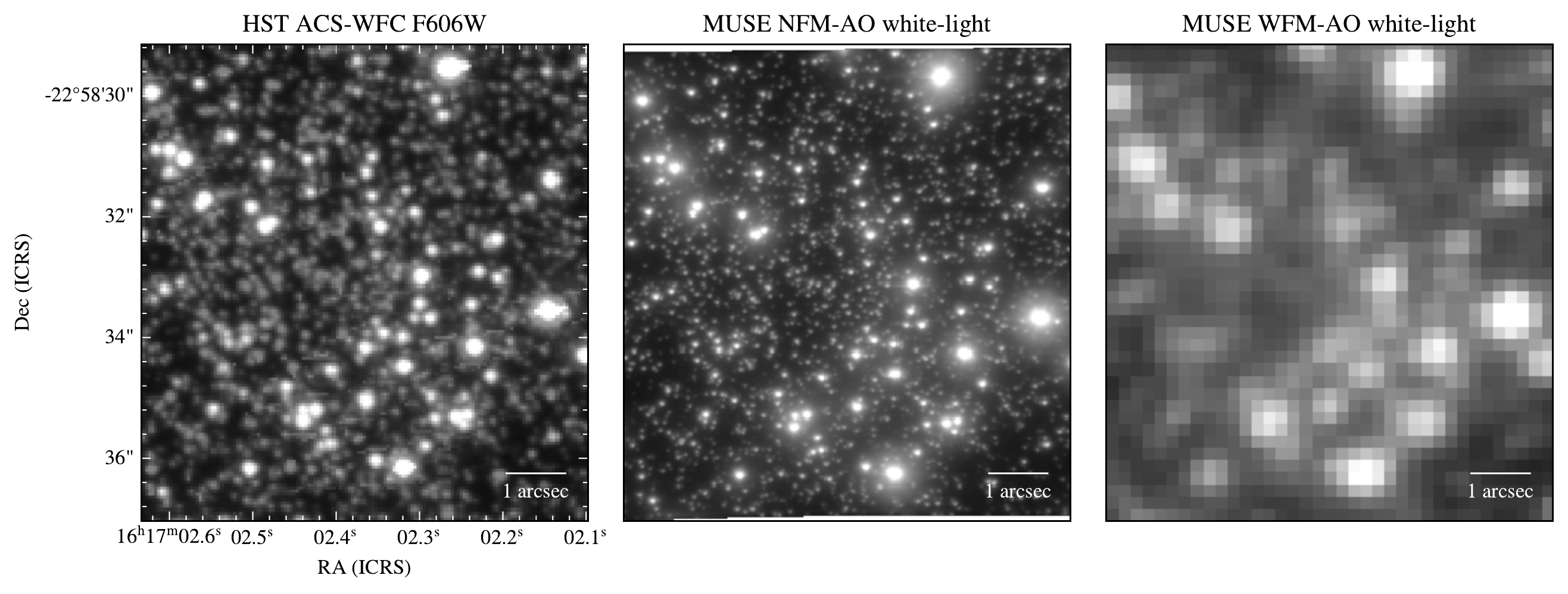}
    \caption{The central 7.5~arcsec by 7.5~arcsec of \ngc{6093} seen with different instruments. Left: the \textit{HST ACS/WFC F606W} image \citep{sarajedini_acs_2007, anderson_acs_2008}, centre: the MUSE NFM white-light image, right: the MUSE WFM-AO white-light image created from the observation with the best seeing (0.4~arcsec). At the cluster distance of 10~kpc, a scale of 1~arcsec corresponds to 0.05~pc. The most complete catalog of this region contains about 1500 stars.}
    \label{fig:nfm_wfm_hst}
\end{figure*}
\label{sect:observations}

This study makes use of spectroscopic data taken with the Multi Unit Spectroscopic Explorer \cite[MUSE,][]{bacon_muse_2010} at the Very Large Telescope as part of the GTO programme `A stellar census in globular clusters with MUSE' (PI: S. Dreizler, S. Kamann) which is described in \cite{kamann_stellar_2018}.
MUSE is an optical integral-field spectrograph which is in operation since 2014. Since mid-2019, a new instrument mode (narrow-field mode, NFM) is offered to the community which enables observations with laser tomography adaptive optics to achieve a higher spatial resolution.
NFM observations have a higher spatial sampling of 0.025~arcsec in a smaller field of view of 7.5~arcsec by 7.5~arcsec compared to a sampling of 0.2~arcsec in a 1~arcmin by 1~arcmin field of view in wide-field mode (WFM) observations. 
While the spatial properties differ, the spectral range in both instrument modes covers 4750 to 9300~\AA{} at a constant sampling of 1.25~\AA{}.

We use all available MUSE observations of \ngc{6093} taken until February 2020. 
The data analysed here includes the ten WFM observations used in \cite{kamann_peculiar_2020}, four new WFM observations, and two NFM observations of a single pointing located in the cluster centre.
Table~\ref{tab:observations} lists the five different pointings we observed and the respective instrument mode. The positions of the pointings are shown in Fig.~\ref{fig:pointings}. 

The main difference of the NFM observations compared to the WFM ones is that we used four instead of three exposures, we did not apply derotator offsets because the natural guide star is off-axis, and the exposure time was 600~s instead of 200~s for each exposure.
We used the most recent versions (2.6 and 2.8) of the MUSE data reduction pipeline \citep{weilbacher_data_2020} to reduce the additional data compared to \citet{kamann_peculiar_2020}.

Figure~\ref{fig:nfm_wfm_hst} compares the image quality of the \textit{HST} \textit{Advanced Camera for Surveys} (\textit{ACS}) \textit{Wide Field Channel} (\textit{WFC}) observation using the \textit{F606W} filter (\citealt{sarajedini_acs_2007, anderson_acs_2008}, zoomed on the cluster core), a white-light image created from the MUSE NFM observation, and a MUSE WFM white-light image of the same region derived from the adaptive-optics observation with the best seeing (0.4~arcsec). 
Clearly, the MUSE WFM observation suffers heavily from crowding, while both the \textit{ACS} and the MUSE NFM image are less affected. 
As described in detail in Section~\ref{sct:extraction} below, we measure a FWHM of around 40~milliarcseconds (40~mas) in the NFM data, i.e. our spatial resolution is higher than that achieved with \textit{ACS-WFC}.

\section{Extracting and analysing spectra}
\begin{figure}
	\includegraphics[width=\columnwidth]{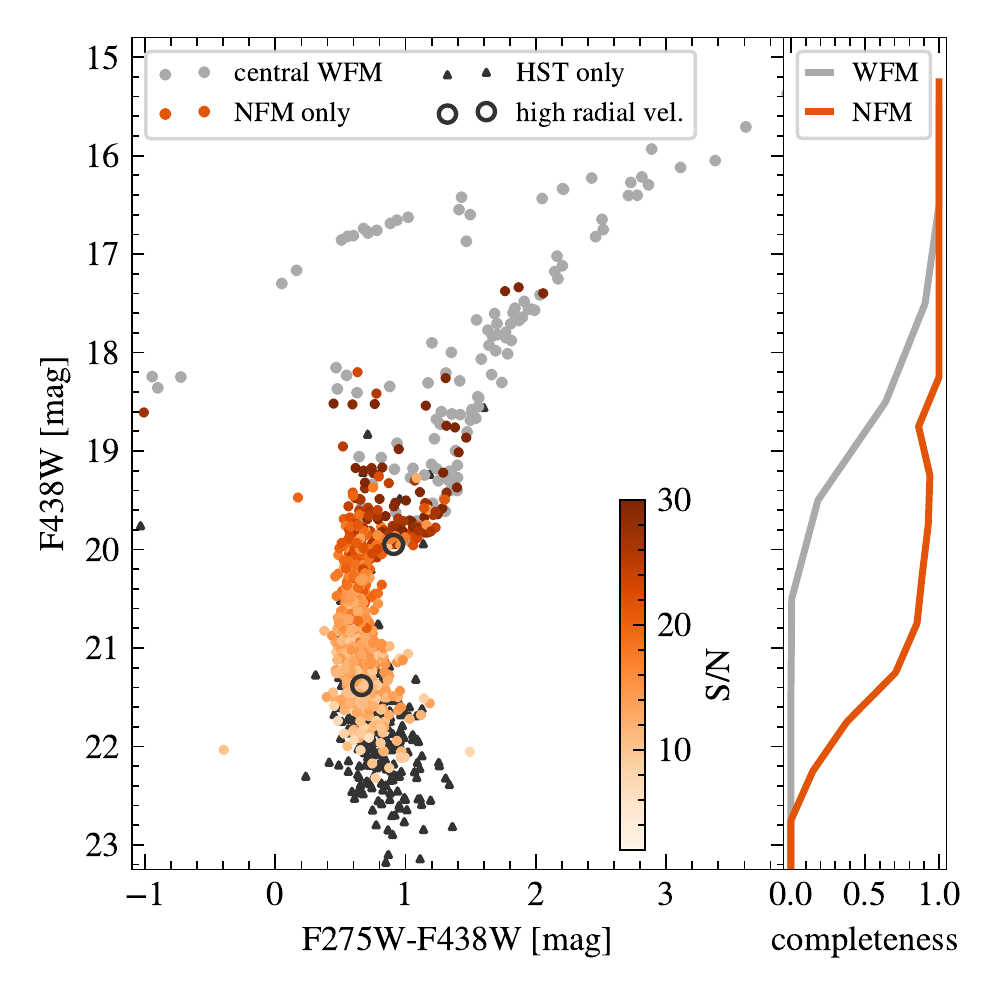}
    \caption{Left panel: CMD of the central region of \ngc{6093} using the photometric catalog of \citet{dalessandro_peculiar_2018}. Grey dots represent stars with spectra extracted from WFM observations while spectra for the red-orange stars could only be obtained from NFM observations. The colour corresponds to the mean S/N ratio from our two NFM observations. Grey triangles are \hst{} sources without extracted spectra and grey open circles are the two stars with a high radial velocity (see Section~\ref{sct:highvrad_stars}).
    Right panel: our spectral extraction completeness relative to the \hst{} catalogue for WFM and NFM observations. In both panels only stars from which we could extract useful radial velocities are taken into account.}
    \label{fig:cmd}
\end{figure}

\label{sect:extraction}

While the spectral extraction and spectral analysis for the WFM observations are identical to the procedure explained in \cite{kamann_peculiar_2020}, we modified them for the NFM data. Here, we describe only these changes.

\subsection{Spectral extraction}
\label{sct:extraction}
We need an initial list of stellar positions and brightnesses at which we extract spectra using \textsc{PampelMuse} \citep{kamann_resolving_2013}.
The \textit{ACS} catalogues compiled by \citet{anderson_acs_2008} have been our standard source for this purpose for most clusters in our survey \citep[see][]{kamann_stellar_2018}, however they do not contain all sources visible in the NFM data of \ngc{6093}. 
\textit{ACS} images of the cluster core taken in the High Resolution Channel are available and we use the catalogue derived by \citet{dalessandro_peculiar_2018} from these data to improve the source catalogue. 
After matching stars included in both catalogues using their positions, the final merged catalogue contains 1500 unique stars in the region covered by our NFM observations. 
However, by visually comparing the combined catalogue with the NFM white-light image, we estimate that 10 to 20~per~cent of all sources visible in the NFM observation still do not have a catalogue entry and thus cannot be extracted. 
Since these missing stars are all faint, we do not expect that we could obtain useful spectral fits and stellar parameters for them, even if a more complete catalogue was available. We also expect their contribution to the spectra of other stars to be negligible. 

\textsc{PampelMuse} reconstructs the shape of the instrumental point spread function (PSF) in a datacube in order to extract spectra. 
While the PSF shape is the well known Moffat profile for MUSE WFM observations, it is more complicated in NFM observations.
We implemented the PSF model \textsc{Maoppy} presented in \cite{fetick_physics-based_2019} in \textsc{PampelMuse} to improve the source extraction. 
The model is designed for the situation typically faced in adaptive optics, where the PSF consists of a coherent core near the diffraction limit surrounded by a seeing-limited halo, the latter being the result of atmospheric turbulence with spatial frequencies uncorrected for by the deformable mirror. 
In the case of \textsc{Maoppy}, the diffraction-limited part of the PSF is adapted to the instrument in use, whereas the atmospheric residuals in the core are modelled as a Moffat function and the uncorrected spatial frequencies are modelled with the Kolmogorov turbulence model, which includes the Fried parameter $r_0$ to scale the turbulence strength.
Figure~\ref{fig:psf_residuals} shows the residuals after PSF and sky subtraction relative to the original data in three different wavelength ranges for the \textsc{Maoppy} model and a combination of a Moffat profile with a Gaussian core.
\changed{As the residuals shown in Fig.~\ref{fig:psf_residuals} slightly increase with increasing wavelength, we suspect that the atmospheric diffraction is not completely corrected. We will learn more about the peculiarities of NFM data after reducing more observations.}
The FWHM of the PSF in our observations decreases from 40~mas in the blue part of the spectrum to 30~mas in the red part.
The Strehl ratio is about 5~per~cent at 650~nm and increases to a maximum of 10~per~cent at 900~nm. 

We further noticed that the catalogue positions of a significant fraction of the resolved stars were not accurate enough at the high spatial resolution offered by the NFM. This became evident when we observed significant bipolar fit residuals around the centroids of the affected stars. 
Recall that by default, \textsc{PampelMuse} predicts the positions of the stars in the MUSE data via a global coordinate transformation from the reference catalogue. The origin of these inaccuracies could be physical (e.g., due to proper motions) or instrumental (e.g. caused by residual errors in the astrometric calibration performed by the data reduction pipeline). 
In order to account for these offset, a new feature was added to \textsc{PampelMuse}, which allows the user to determine individual offsets $\delta x$ and $\delta y$ for each star relative to the NFM positions predicted by the global coordinate transformation. In the case of \ngc{6093}, typical offsets of $0.4$~spatial pixels, corresponding to 10~mas, were applied.
For comparison, a star with velocity of 10~\kms{} in the plane of the sky at a distance of 10~kpc has moved about 3~mas since the \hst{} observations in 2006~\citep{anderson_acs_2008}.
Further details on the improvements implemented in \textsc{PampelMuse} in order to deal with NFM data will be presented in a forthcoming paper.

\subsection{Spectral analysis}
To analyse the extracted spectra, we use our well-tested fitting pipeline described in \citet{husser_muse_2016}.
Since the adaptive-optics correction works better in the red than in the blue part of the spectrum, the spectral noise is higher in the blue part.
To take care of this systematic difference, the full-spectrum fit uses uncertainties derived from the variance extension of the datacube as weights for spectra extracted from NFM observations. 

\subsection{Improvements due to NFM observations}

Compared to the previous study of \ngc{6093} \citep{kamann_peculiar_2020}, we obtain spectra of more than a thousand new stars in the central 7.5~arcsec by 7.5~arcsec from our NFM observations. Here, we only take into account stars with ``useful'' spectra which are spectra with a S/N above five, a radial velocity reliability of at least 90~per~cent (see Section~\ref{sct:reliable}) and a \texttt{MagAccuracy} of more than 80~per~cent \citep[see definition in ][]{kamann_stellar_2018}. We later use radial velocities resulting only from spectra that fulfil these criteria.
Figure~\ref{fig:cmd} shows a colour-magnitude diagram of stars inside the NFM region and indicates whether we have observed useful spectra with the WFM, the NFM, or if we did not get a useful spectrum.
We analysed the spectra of 185 stars from the WFM observations of the cluster centre, 176 of them have spectra with a mean S/N of at least five, and 121 have a mean S/N of at least ten.
With the NFM observations, we gain 932 stars that had no useful analysis result from previous observations. Of those 932 stars, 891 have spectra with a mean S/N of at least ten.
The total number of stars with useful spectra inside the NFM footprint is 1072. This implies an overall completeness of 71~per~cent compared to the combined photometric catalog of \cite{anderson_acs_2008} and \cite{dalessandro_peculiar_2018} which contains 1501 stars in the region covered by our NFM observations. 

\subsection{Filtering data for reliability and binarity}
\label{sct:reliable}
As described in \cite{kamann_stellar_2018}, we estimate whether a star is a member of the cluster or a foreground star based on a model of the Galactic stellar population \citep{robin_synthetic_2003} in the direction of the cluster. Depending on its mean radial velocity and metallicity, each star is assigned a membership probability.

Regardless of whether a spectrum is extracted from a NFM or WFM observation, its radial velocity derived from the spectral fit is assigned a reliability. 
This reliability is defined in \citet[Section~3.2]{giesers_stellar_2019} and it depends on the following properties:
the S/N ratio of the respective spectrum, the quality of a cross-correlation with a suitable template model spectrum, the difference of the radial velocity derived from the cross-correlation and the full-spectrum fit, plausible uncertainties of the radial velocity from the cross-correlation and full-spectrum fit, and agreement between the fitted radial velocity and the mean cluster radial velocity. 
In this study, we include all radial velocities resulting from a fit with a reliability of more than 90~per~cent.

When multiple radial velocities are available for a given star, we use the method described in \cite{giesers_stellar_2019} to compute the probability $p$ that they show temporal variability. 
Since the dynamical model we use is not able to take binary stars into account, we follow \citet{kamann_peculiar_2020} and only include stars with $p < 80$~per~cent, removing 169 stars from further analysis (about 2~per~cent of our final sample of stars).
We average all reliable radial velocities of the same star weighted by their uncertainties to obtain a mean radial velocity. 
After filtering, we check the consistency of WMF and NFM observations by computing the mean radial velocity per star separately for WFM and NFM observations. For stars which were observed in both modes, the weighted mean difference of the WFM and NFM velocities is $-0.03 \pm 0.77$~\kms{}.
The number of stars in our analysis after filtering is 9720.
In the further analysis, we use the measured radial velocities after subtracting the mean radial velocity of about 9.7~\kms{}.

This procedure removes two interesting stars very close to the \citet{goldsbury_acs_2010} cluster centre (less than about 1~arcsec) with high radial velocities of $(88 \pm 7)$ and $(101\pm 4)$~\kms{} relative to the Solar System barycentre, respectively, and about 10~\kms{} less relative to the cluster. 
It is plausible that both stars are cluster members and not foreground stars because of their low metallicities (both have $\text{[M/H]} \approx -2$) and their positions in the colour-magnitude diagram (on the main sequence and subgiant branch, see~Fig.~\ref{fig:cmd}). We discuss these stars in Section~\ref{sct:highvrad_stars}.

\subsection{External kinematic data}
To increase the coverage of the outer parts of the cluster, we also include the radial velocities of \cite{baumgardt_catalogue_2018}. 
Stars that appear in both data sets have compatible velocities as shown by \citet{kamann_peculiar_2020}.
Before combining these data with ours, we subtract the mean radial velocity from each individual velocity.
\cite{baumgardt_catalogue_2018} provide a probability \texttt{P\_single} indicating whether a given star is a single star. We use this information to exclude stars with ${\rm \texttt{P\_single}}$ less than 20~per~cent and use the remaining 230 stars in our analysis. Their radial distances to the cluster centre are between 20~arcsec and 20~arcmin with a median of 3~arcmin.
The total number of stars in our analysis is 9950.
There are no proper motions available for the central regions of \ngc{6093} because it is not part of the \hst{} proper motion programme \citep{bellini_hubble_2014}.

\section{Jeans model}
\label{sect:model_upper}
\subsection{Description}
\label{sect:model}

We use the axisymmetric Jeans model code \textsc{cjam} of \cite{watkins_discrete_2013} which is based on \textsc{jam} \citep{cappellari_measuring_2008}. 
\textsc{cjam} was also used in the previous analysis of \cite{kamann_peculiar_2020}. 
These Jeans models include the effects of anisotropy and rotation to calculate the first and second moments of the velocity distribution at a given point.
We compared the model predictions to our data using the same maximum-likelihood approach as in \citet{kamann_peculiar_2020}.
The advantage of this approach is that it does not bin the velocities but works with all individual data points. While this increases the computational complexity, it removes the subjective binning step from the analysis.

Since we share the basic model and analyse data of the same globular cluster, we adopt several parameters from the study of \citet{kamann_peculiar_2020}.
In particular, we assume an axis ratio of $q = 0.9$ for the cluster elongation.
We also assume isotropy, i.e. the velocity dispersion along the line of sight and those tangential to it are identical ($\beta = 0$).
This assumption is justified by the core relaxation time of $10^{7.78}$~years \citep{harris_catalog_1996} which suggests isotropy at the core radius \citep[][Section 5.4]{watkins_hubble_2015}. 

Compared to the previous study, the model differs in the following aspects:
\begin{itemize}
    \item The coordinates for the cluster centre are no longer fixed to the literature values. Instead, we use two new parameters, $\Delta x$ and $\Delta y$, with Gaussian priors centred on the photometric centre of \citet{goldsbury_acs_2010} to account for  uncertainties in the determination of the cluster centre.
    \item We compute a two-dimensional grid of 960 profiles in the form of Multi-Gaussian expansions that cover centre offsets between $-3$~arcsec and $+3$~arcsec relative to the \citet{goldsbury_acs_2010} photometric centre and use the MGE of the closest grid point to compute the Jeans model during each MCMC step. To construct these profiles we used the \hst{} photometry of \citet{dalessandro_peculiar_2018} complemented by the Gaia data of \citet{de_boer_globular_2019} as described in \citet{kamann_peculiar_2020}, Section~4.1.
    \item As suggested by \cite{hogg_data_2018}, the rotation field is now parametrized by two vector components, $\kappa_x$ and $\kappa_y$, instead of its amplitude and an angle. 
    While \citet{kamann_peculiar_2020} determined the position angle before the main MCMC run, we vary these parameters together with all remaining ones during the MCMC run. As the Jeans code assumes the semi-major axis to be aligned with the x-axis of the coordinate system, we rotate our data by the current position angle estimate prior to calculating a new model.
    \item We include a potential for an IMBH parametrized by its mass. \textsc{jam} and \textsc{cjam} implement this by adding an additional Gaussian component to the MGE. At a mass of zero, the total potential is equal to the potential without a central IMBH. 
    We take a uniform probability distribution between 0 and $15{,}000~\msol{}$ as a prior for the IMBH mass.
    We follow a comment in \citet{cappellari_measuring_2008} and fix the standard deviation of the corresponding MGE component to $\sigma_\text{PSF}/3 \approx 25$~mas.
    
\end{itemize}

For a given set of parameters, the Jeans code predicts the mean velocity and the velocity dispersion at the location of each star from our kinematic data set. 
As usual for MCMC approaches, we calculate parameter distributions by maximizing the product of the prior and the likelihood function, which is a Gaussian centered on the difference of measured velocities and the model prediction at that position \citep[][Eq. 12]{watkins_discrete_2013}.
Table~\ref{tab:priors} lists the eleven parameters and their respective priors.
We use the affine invariant ensemble MCMC sampler \textsc{emcee} \citep{foreman-mackey_emcee_2013}. We use 192 walkers and a chain length of 2000 after burn-in. 

\subsection{Results}
\begin{figure}
	\includegraphics[width=\columnwidth]{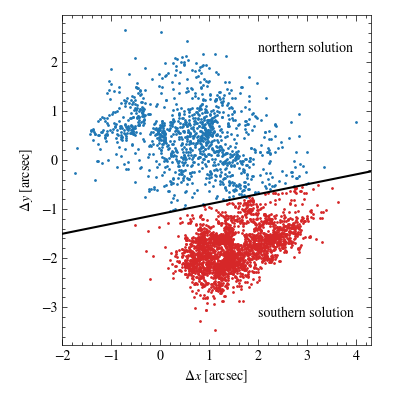}
    \caption{\changed{The distribution of centre offsets $\Delta x$ and $\Delta y$ is bimodal. We use the line $Y = 0.2 \Delta x - 1\farcs{}1$ to separate the samples into a northern solution with $\Delta y > Y$ and a southern solution with $\Delta y < Y.$}}
    \label{fig:clustered_xy}
\end{figure}
The parameter distributions that we generate with the MCMC sampler (see Fig.~\ref{fig:corner}) are bimodal in the centre offsets $\Delta x$ and $\Delta y$. \changed{As shown in Fig.~\ref{fig:clustered_xy},} we separate the samples along the line $Y = 0.2 \Delta x - 1\farcs{}1$ in the $\Delta x \Delta y$-plane. About one third of all samples have $\Delta y \geq Y$ and we name them `northern solution', we name the other two thirds with $\Delta y < Y$ `southern solution'. 
\changed{The four circular gaps visible in the southern solution are each centered on a grid point of the MGE grid. The MGE at these locations have one component fewer than the surrounding ones and while they fit the surface brightness profile, they apparently cause a decrease in the likelihood when used as input for the Jeans model to fit the velocities.}
We present the overall parameter distribution resulting from the MCMC sampling in this section when there is no difference between the northern and the southern solutions and comment on the difference if there is any. 

The profile of the mass-to-light ratio $\Upsilon(r)$ has a minimum at 
$r_\Upsilon = 1.4^{+0.9}_{-0.8} \arcmin{}.$
We find a mean value of $\Upsilon = 1.87 \pm 0.15 \, \msol \, \lsol{}^{-1}$ for the cluster which is consistent with the value of $\Upsilon = 1.72 \pm 0.20 \, \msol \, \lsol{}^{-1}$  determined in \citet{kamann_peculiar_2020} and also with 
$\Upsilon = 1.93 \pm 0.12 \, \msol \, \lsol{}^{-1}$ from \citet{baumgardt_absolute_2020}. 
We plot profiles of the mass-to-light ratio $\Upsilon(r)$ using random samples drawn from our chain in Fig.~\ref{fig:mlr}. 
The total cluster mass is $(3.0_{-0.3}^{+0.2}) \times 10^5 ~\msol{}.$

For the median intrinsic flattening we find $\bar q = 0.86^{+0.03}_{-0.05}$ which is also consistent with $\bar q = 0.83 \pm 0.06$ found by \citet{kamann_peculiar_2020}.
This corresponds to an inclination of $60\degr{} \pm 15 \degr{}.$

The histograms of all eleven free parameters of the fitted Jeans model and their pairwise correlations are shown in Figure~\ref{fig:corner}. 

\subsubsection{Dispersion and rotation profiles}
\begin{figure*}
	\includegraphics[width=2\columnwidth]{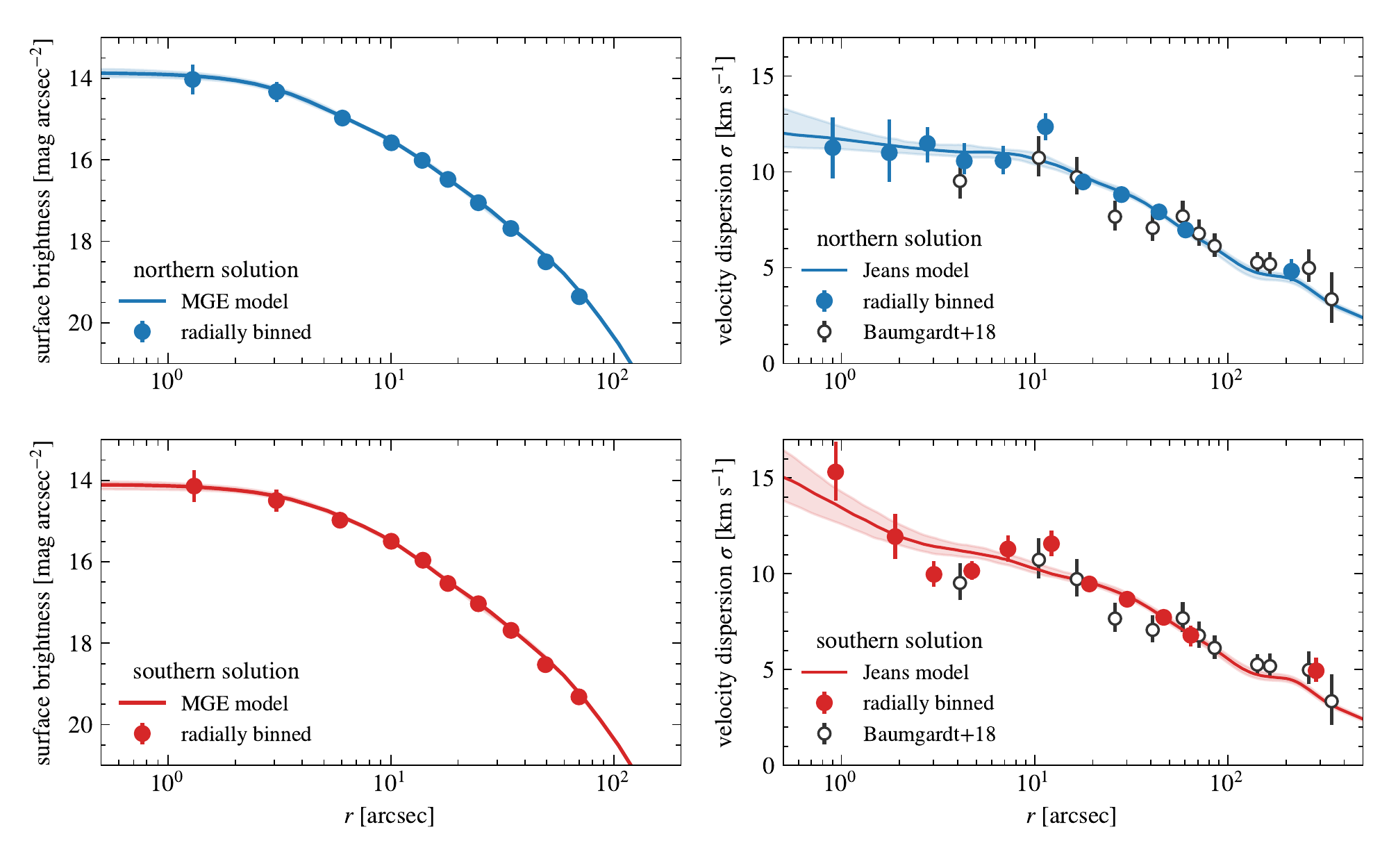}
    \caption{\changed{Radial profiles of the surface brightness (left column) and the velocity dispersion (right column). In the left column, filled circles show the profile derived using the photometry presented in \citet{dalessandro_peculiar_2018}, while the solid line shows our best-fit MGE model. In the right column, filled circles indicate the dispersion obtained directly from the MUSE data, whereas solid lines indicate the predictions from the Jeans models. In all panels, errorbars and shaded areas represent the 14th and 86th percentiles. The top row includes only MCMC samples from the northern solution, the bottom row only from the southern solution. The reference data from \citet{baumgardt_catalogue_2018} is plotted as published, i.e. without taking into account any centre offset.}}
    \label{fig:dispersion}
\end{figure*}

\begin{figure}
	\includegraphics[width=\columnwidth]{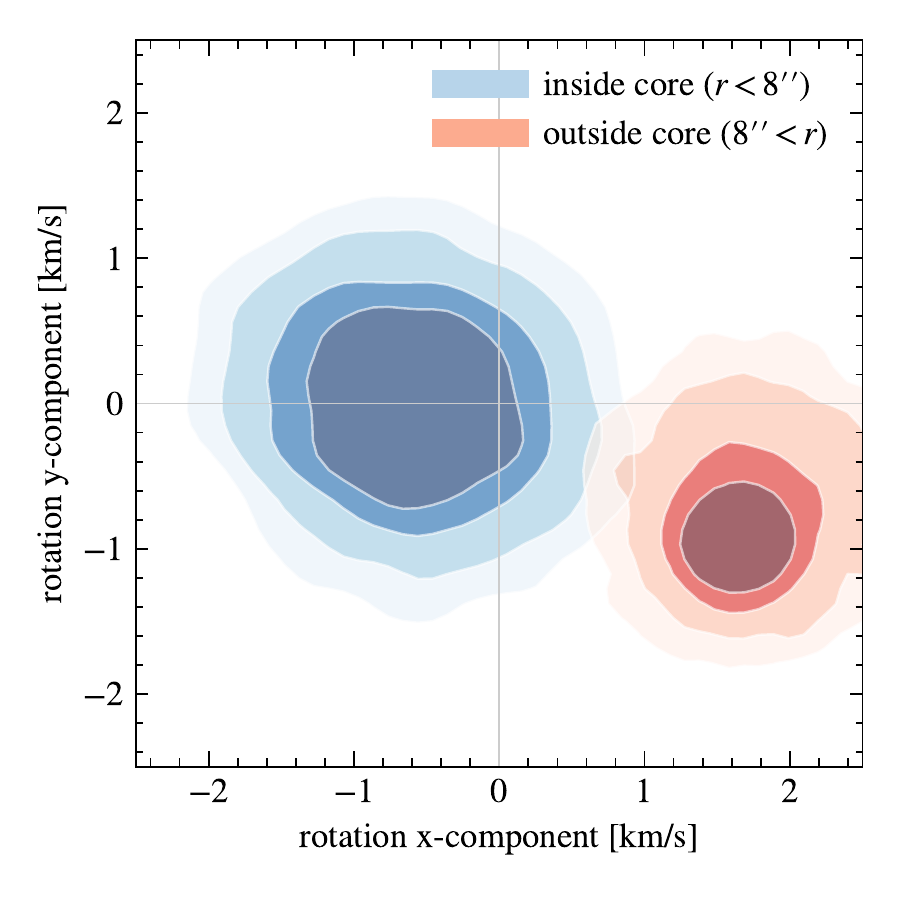}
    \caption{Joint distribution of the estimated x- and y-component of the rotation velocity. Each set of contours contains 50, 68, 87, and 95~per~cent of all samples (from inner- to outermost) belonging to the core region (blue) or the outer region (red) of \ngc{6093}. The outer region is rotating, but it is not clear if the cluster core ($r < 8$~arcsec) is also rotating.}
    \label{fig:2drotation}
\end{figure}

To compare if the complex kinematic model actually reproduces our data, we plot the velocity dispersion profiles computed from the parameters in the chain with our data in Fig.~\ref{fig:dispersion}. \changed{This figure also shows a comparison of the observed surface brightness profile and the fitted MGE models.}

We binned the data radially and applied a much simpler kinematic model to each bin which determines a constant velocity dispersion and the components $v_x$ and $v_y$ of the rotational vector. 
In this model, the predicted radial velocity follows a Gaussian probability distribution with dispersion $\sigma$ and a mean that depends on the model parameters $v_\text{sys}$, $v_x$, and $v_y$ and on the position angle $\theta$ of a star according to
\begin{equation}
\label{eq:simple_model}
v(\theta) = v_\text{sys} + \sqrt{v_x^2+v_y^2} \, \sin[ \theta-\arctan(v_x/v_y)],    
\end{equation}
where $\theta - \arctan(v_x/v_y)$ is the angular distance to the rotation axis.
To account for the uncertainty of the cluster centre coordinates, we repeat the binning for 250 potential centres drawn from the MCMC chain.
The values from this simple model and the Jeans model agree with each other. 
Furthermore, Fig.~\ref{fig:dispersion} shows the velocity dispersions used in the analysis of \ngc{6093} by \citet{baumgardt_catalogue_2018} which also agree with our values.
The Jeans model has a central velocity dispersion (at radius of 1~arcsec) of $12.8 \pm 1.3$~\kms{}, which is about 2~\kms{} more than the value of $10.5 \pm 0.5$~\kms{} from \citet{baumgardt_catalogue_2018}. The northern solution has a lower central velocity dispersion of $11.7 \pm 0.6$~\kms{} and the southern solution has a central velocity dispersion of $13.4 \pm 0.8$~\kms{}.
To check if the higher values for the central velocity dispersion of the southern solution is an artefact caused by a few stars with unusual radial velocities, we plotted the radial velocity of individual stars close to the respective centres as a function of the centre distance. We did not find any outliers in these plots. Instead, the overall scatter in the radial velocities is larger around the centre of the southern solution.
\changed{
To further check how reliable the increased velocity dispersion around the southern centre is, we compute the biweight scale of the radial velocities, a robust measure of the standard deviation \citep[see][]{beers_measures_1990} for the $n$ nearest neighbours of each star in the NFM field of view ($n = 20, 40, 100$). This method also shows an increase in the velocity dispersion around the southern centre but not around the northern one.
}

We find an overall rotation angle of $66\degr{} \pm 6\degr{}$, consistent with the angle of $60\degr{} \pm 3\degr{}$ found by \citet{kamann_peculiar_2020}.
Figure~\ref{fig:rotation} shows the radial profile of the rotation amplitude derived from the final MCMC chain. That figure also shows the rotation amplitude computed using the same radial bins and model as for the dispersion plot described above. 
The radial rotation profile follows the one from \citet{kamann_peculiar_2020} as expected. The binned data show a very large uncertainty towards the cluster centre because the bin radii are similar to the uncertainty in the centre coordinates. 

As a consistency check, we also compute the biweight location of the radial velocities, a robust measure for the mean \citep{beers_measures_1990}, for $n=250$ nearest neighbours of each star in the central $100\arcsec{} \times 100\arcsec{}$. We calculate the rotational parameters from these velocities by treating them as ordinary velocities in our simple model (Eq.~\ref{eq:simple_model}). 
\changedtwo{As demonstrated in Fig.~\ref{fig:rotation}, the} rotational amplitude and angle derived in this way have a similar radial profile as the ones directly computed from the radial velocities.

\begin{figure}
	\includegraphics[width=\columnwidth]{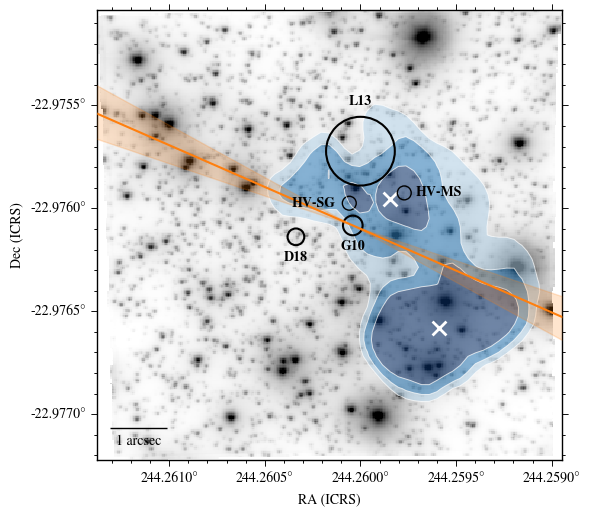}
    \caption{Contour plot of our estimate of the kinematic centre of \ngc{6093}. The blue-shaded regions cumulatively contain 68, 87, and 95~per~cent of all samples (from inner- to outermost). The two white crosses are the centres of the northern and the southern solution. G10 denotes the centre found in \citet{goldsbury_acs_2010}, L13 is from \citet{lutzgendorf_limits_2013}, and D18 is the centre of \citet{dalessandro_peculiar_2018}. The smaller circles mark the potential central high-velocity main-sequence (HV-MS) and subgiant (HV-SG) stars. \changed{The orange line indicates the rotation axis of the whole cluster and its uncertainty computed from the 16th and 84th percentiles, using G10 as the reference centre.}}
    \label{fig:center}
\end{figure}

We had a closer look at whether the core of M80 rotates. While such a central rotation component is expected to be short-lived because of the effects of two-body relaxation, evidence for a decoupled core has for example been reported in the core collapse cluster \ngc{7078} \citep{van_den_bosch_dynamical_2006,usher_muse_2021}.
The central rotation curve strongly depends on the position of the cluster centre since we use the radial distance to the centre for binning. 
To account for the uncertainty in the position of the cluster centre, we draw 250 pairs of the centre position offsets $\Delta x$ and $\Delta y$ from our final MCMC chain.
Figure~\ref{fig:2drotation} shows the joint distribution of the x- and y-component of the rotation velocity for different radial bins.
While we find a clear rotation signal of $v_\text{rot} = 1.9 \pm 0.4$~\kms{} in the outer parts of the cluster, similar to the value of $1.97 \pm 0.84$~\kms{} determined by \citet[][Table 1]{sollima_eye_2019}, it is less clear if the cluster core 
\citep[$r_\text{core}\approx8\arcsec{},$][]{harris_catalog_1996,baumgardt_catalogue_2018} is rotating. In this region, the rotation components follow a broad distribution with a median of 0.9~\kms{}, 90~per~cent of all samples are below $v_\text{rot} = 1.6$~\kms{}. 
We note that the central rotation angle is offset from the rotation angle in the outer parts of the cluster by about $150\degr{}\pm 30$\degr{}.

\subsubsection{Position of the cluster centre}
The centres of globular clusters are usually determined using photometric data. 
There are three recent measurements for the cluster centre of \ngc{6093}. \citet{goldsbury_acs_2010} found
$\text{RA} = 16\text{h}\,17\text{m}\,2.41\text{s}$ and
$\text{Dec} = -22\degr{} 58\arcmin{} 33\farcs{}9 $ with an uncertainty of $0\farcs{}2$, while
\citet{dalessandro_peculiar_2018} found
$\text{RA} = 16\text{h}\,17\text{m}\,2.481\text{s}$ and $\text{Dec} = -22\degr{} 58\arcmin{} 34\farcs{}098 $ with an uncertainty of $0\farcs{}17$, and \citet{lutzgendorf_limits_2013} found
$\text{RA} = 16\text{h}\,17\text{m}\,2.4\text{s}$ and $\text{Dec} = -22\degr{} 58\arcmin{} 32\farcs{}6$ with an uncertainty of $0\farcs{}7$.
We compare these centres with the offsets in the centre position as sampled by our MCMC chain by transforming the offsets to equatorial coordinates and plotting their density in Figure~\ref{fig:center}.
The centre of the northern solution is located at 
\begin{equation}
\text{RA} = 16\text{h}\,17\text{m}\,2.36\text{s},~\text{Dec} = -22\degr{} 58\arcmin{} 33\farcs{}4,
\end{equation}
and the centre of the southern solution is located at
\begin{equation}
\text{RA} = 16\text{h}\,17\text{m}\,2.30\text{s},~\text{Dec} = -22\degr{} 58\arcmin{} 35\farcs{}7.
\end{equation}
As these centres are determined from a dynamical model and to distinguish it from the photometric centres, we call them dynamical centres.
The contours around the dynamical centres in Fig.~\ref{fig:center} show the uncertainties which are larger than those for the photometric centres.
While the centre of the northern solution is close to the known photometric centres, especially to the ones from \citet{goldsbury_acs_2010} and \citet{lutzgendorf_limits_2013}, the southern centre is located at a distance of about 2.4~arcsec to the south-west of the northern centre.

\subsubsection{Limits on the IMBH mass}

\begin{figure}
\centering
	\includegraphics[width=\columnwidth]{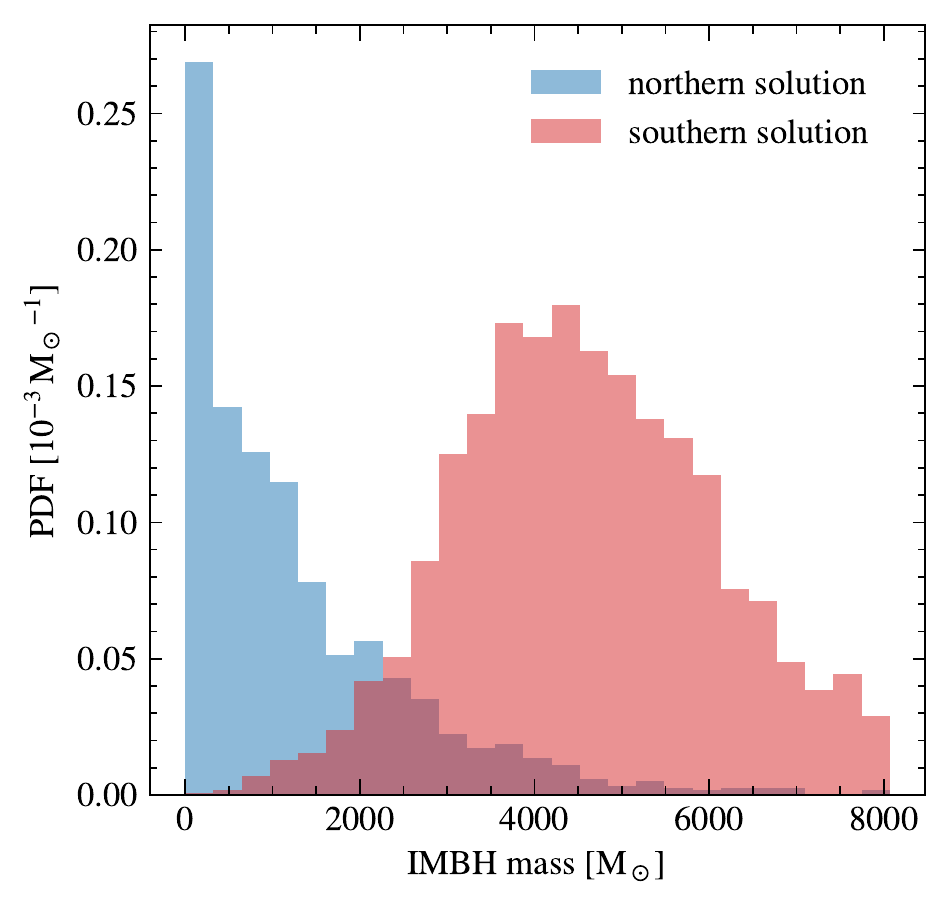}
    \caption{Distribution of the IMBH masses obtained via Jeans modelling around different centres.
    }
    \label{fig:imbh_centre}
\end{figure}

The posterior distribution of the mass has a peak at masses below $1000~\msol{}$ and another peak at about $5000~\msol{}$. 
The uncertain position of the centre has a large impact on the shape and moments of the distribution of the IMBH mass, as can be seen from Fig.~\ref{fig:imbh_centre}. 
Here, we compare the IMBH mass distributions of the northern and the southern solutions.
The 90-per-cent upper limits on the IMBH mass (rounded to the nearest $100~\msol{}$) are $3000~\msol{}$ and $6800~\msol{}$, respectively.
Given a peaked shape of the IMBH mass distributions for the southern centre, we also report a median IMBH mass and an $1\sigma$ uncertainty of $4600^{+1700}_{-1400}~\msol{}$ around this centre.

\section{N-body models}
\label{sect:nbody}
\begin{figure*}
	\includegraphics[width=2\columnwidth]{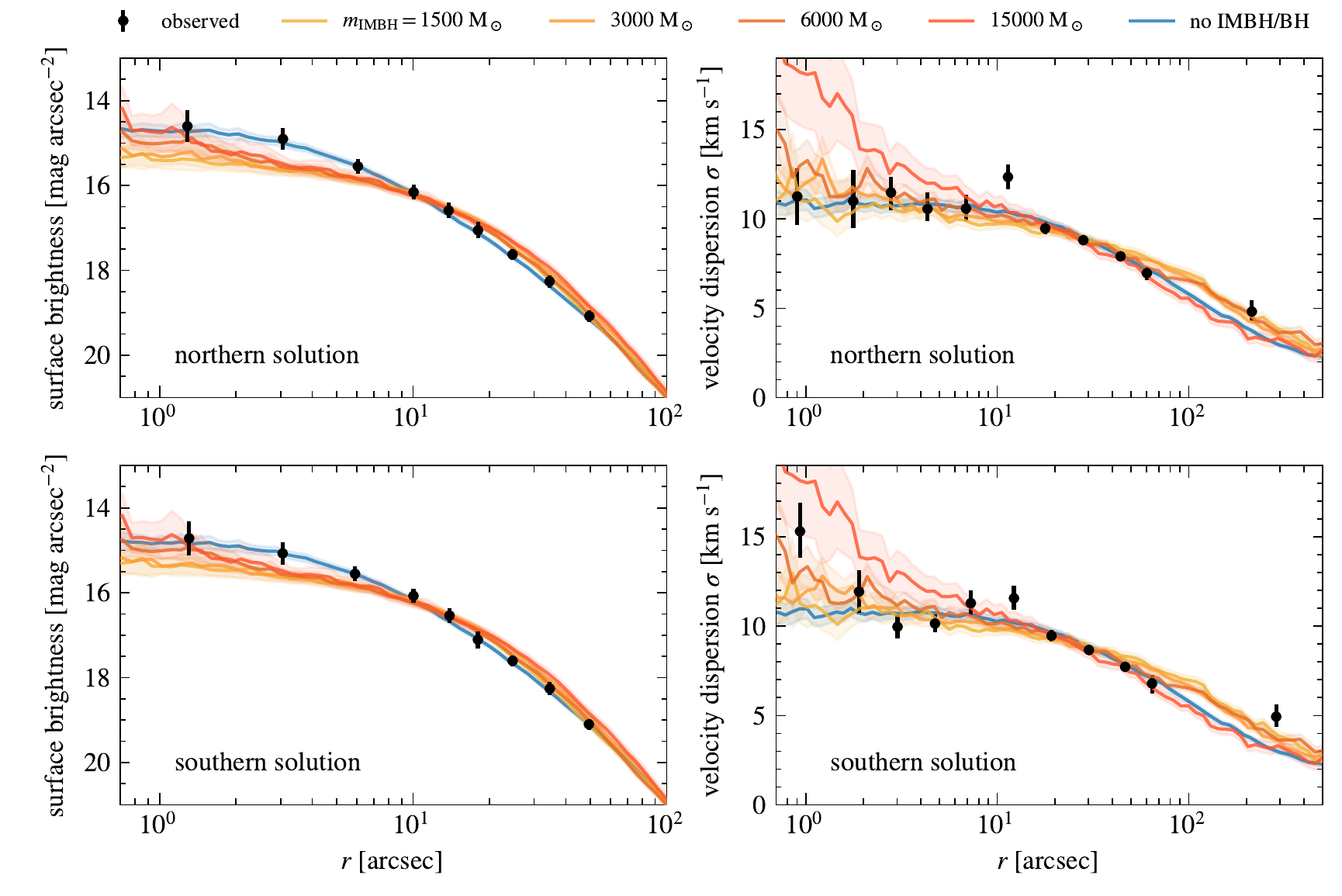}
    \caption{\changed{Comparison of the observed surface brightness and the observed velocity dispersion computed from radial bins around the northern and southern centre, respectively, to $N$-body models with an IMBH. See Tables~\ref{tab:fits_north} and \ref{tab:fits_south} for more details on the models.}}
    \label{fig:nbody_imbh}
\end{figure*}

\begin{figure*}
	\includegraphics[width=2\columnwidth]{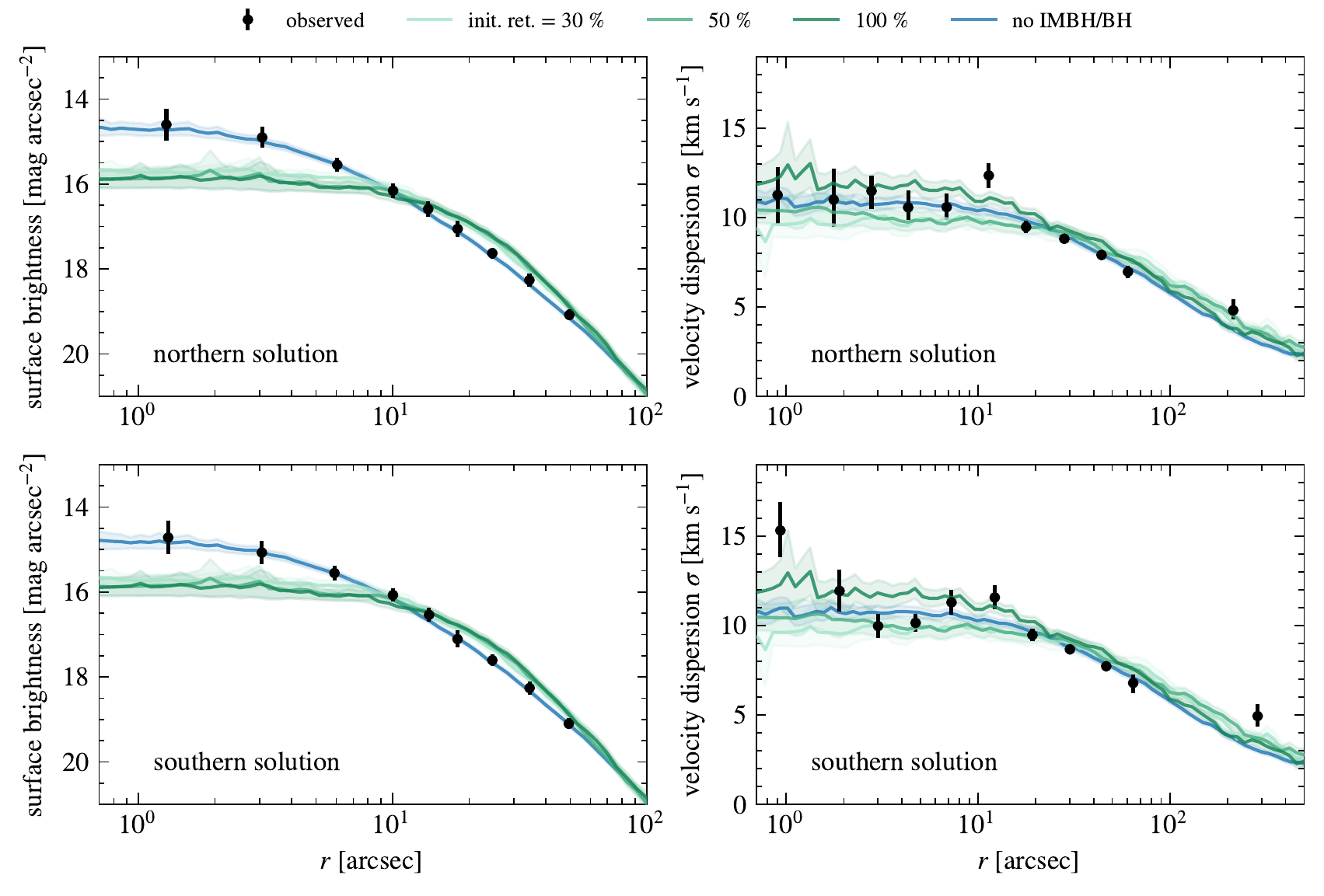}
    \caption{\changed{Comparison of the observed surface brightness and the observed velocity dispersion computed from radial bins around the northern and southern centre, respectively, to $N$-body models with stellar-mass BHs. The $N$-body models have different initial retention fractions for stellar-mass BHs, see Tables~\ref{tab:fits_north} and \ref{tab:fits_south} for more details on the models.}}
    \label{fig:nbody_mbh}
\end{figure*}

In order to further investigate if the observed velocity dispersion of \ngc{6093} requires the presence of an IMBH, we fit a grid of
$N$-body models against the observed surface brightness and velocity dispersion profiles. In particular we use the grid of IMBH models presented
by \citet{baumgardt_n-body_2017} and the grid of models with varying stellar-mass back hole retention fractions presented in \citet{baumgardt_no_2019}.
\citet{baumgardt_n-body_2017} and \citet{baumgardt_no_2019} have run a grid of about 3000 $N$-body simulations of star clusters containing $N=100{,}000$
or $N=200{,}000$ stars using \textsc{nbody6} \citep{aarseth_nbody1_1999,nitadori_accelerating_2012}, varying the initial density profile and half-mass radius, the
initial mass function and the mass fraction of an intermediate mass black hole in the clusters.
Figure \ref{fig:nbody_imbh} compares the best-fitting models without an IMBH and with IMBHs containing 0.5, 1.0, \changedtwo{2.0 and 5.0~per~cent} of the final cluster mass \changedtwo{of $3 \times 10^5~\msol{}$} with the observed velocity dispersion profile and the observed surface brightness profile for the northern and southern centre solutions (also see Tables~\ref{tab:fits_north} and \ref{tab:fits_south}). 
\changedtwo{Around the southern centre} it can be seen that models with an IMBH lead to somewhat better fits of the velocity dispersion profile, whereas the surface brightness profile is better reproduced by the models without an IMBH. Measuring the error weighted difference between observed and predicted velocity dispersion and dividing by the number of degrees of freedom, \changed{we obtain a reduced $\chi^2_r$ value of 1.54 for the 2-per-cent IMBH model from the fit to the velocity dispersion profile, while a no-IMBH solution gives $\chi^2_r=2.25$. Including the fit to the surface brightness profile as well, the reduced $\chi^2_r$ value is 1.16 for the 2-per-cent IMBH model and 1.51 for the  0.5-per-cent IMBH model while the no-IMBH model has $\chi^2_r=1.24$.}
Hence the $N$-body models confirm the results of the Jeans modelling that an IMBH of a mass of around 6000~$\msol{}$ provides an acceptable fit for \ngc{6093} if the southern solution is adopted as the density centre. A no-IMBH model does however provide a good fit around this centre \changedtwo{as well.}
Around the northern centre we find similar fit results. \changed{The no-IMBH solution provides the best fit but the IMBH models result only in slightly worse fits and are also acceptable solutions. Again, the surface brightness profile is better reproduced by a model without any IMBH.} Thus, if the northern centre is adopted, an IMBH with a mass of up to about 6000~$\msol{}$ is possible \changedtwo{as well}.

The corresponding fit of models with different retention fractions of stellar-mass black holes is depicted in Fig.~\ref{fig:nbody_mbh}. 
\changed{We obtain rather poor fits to the surface brightness profile for models with retention fractions of stellar mass black holes greater than 30~per~cent.
The  fits become worse the larger the assumed retention fraction of stellar-mass black holes. 
This is particularly striking for the surface brightness profile, as all models with significant retention fractions produce a core that is larger than what is observed, in agreement with the prediction that clusters harbouring a considerable number of black holes should have large core radii \citep[][]{arca_sedda_mocca-survey_2018}.}
We therefore conclude that the initial retention fraction of black holes in \ngc{6093} was low or that nearly all stellar-mass black holes have been ejected from this cluster.

\section{Discovery of two stars with high radial velocity}
\label{sct:highvrad_stars}
We find two stars very close to the \cite{goldsbury_acs_2010} cluster centre (with a distance less than about 1~arcsec, see Fig.~\ref{fig:center}) with a high radial velocity relative to the Solar System barycentre: a main-sequence star with $v = (88 \pm 7)$~\kms{} at a projected distance of 1.1~arcsec ($\text{RA} = 16\text{h}\,17\text{m}\,2.345\text{s}$, $\text{Dec} = -22\degr{} 58\arcmin{} 33\farcs{}33$)
and a subgiant star with $v = (101\pm 4)$~\kms{} at a projected distance of 0.4~arcsec ($\text{RA} = 16\text{h}\,17\text{m}\,2.414\text{s}$, $\text{Dec} = -22\degr{} 58\arcmin{} 33\farcs{}51$).
While we have two spectra of the main-sequence star and three spectra of the subgiant, we could only derive one and two useful radial velocities respectively due to the low signal-to-noise of the remaining spectra.
The radial velocity of both stars is well above the central escape velocity of \ngc{6093} \citep[41.4~\kms{},][]{baumgardt_catalogue_2018}.

High-velocity stars have been reported in three other clusters: in \ngc{2808} \citep{lutzgendorf_high-velocity_2012}, in M3 \citep{gunn_dynamical_1979}, and in 47~Tuc \citep{meylan_two_1991}. In these cases, two stars were detected with a radial velocity a few times the velocity dispersion above or below the mean cluster velocity but below the central escape velocity of the respective cluster.

The high radial velocity of these two stars can be explained in several ways:
\paragraph*{Foreground star} One or both stars could be foreground stars instead of cluster members. 
Using the Besan\c{c}on model \citep{robin_synthetic_2003} of the region close to \ngc{6093}, we estimate that about 0.7~per~cent of all foreground stars have $-2.25 < \text{[Fe/H]} < -1.75$, $|v| > 80$~\kms{} and a V brightness above 21~mag. As we find $n=25$ non-member stars in the central 6~arcsec, the probability of having one high-velocity non-member star in this region is 16~per~cent, and it is 5~per~cent for the central 1.5~arcsec. The probability that two or more such stars are present is about 1~per~cent and 0.04~per~cent, respectively. 
\changedtwo{
In order to further investigate the membership of the fast-moving stars, we analysed two sets of HST observations, one obtained in 2006 (HST proposal ID 10775, PI Sarajedini) and one taken in 2012 (HST proposal ID 12605, PI Piotto). We analysed each data set using \textsc{Dolphot} \citep{dolphin_wfpc2_2000,dolphin_dolphot_2016} and transformed the HST coordinates into the ICRS reference frame by cross-matching the stellar coordinates against those of the Gaia EDR3 catalogue \citep{gaia_collaboration_gaia_2021}. For each data set, we performed an epoch transformation of the Gaia coordinates back to the epoch when the HST observations were taken, using the measured Gaia proper motions of the stars. Comparison of two sets of stellar coordinates obtained from the cross-matching then allows us to calculate the absolute proper motions of the stars. We obtain a mean cluster motion of ($\mu_\alpha*, \mu_\delta$) = ($-3.03 \pm 0.05, -5.61 \pm 0.05$) mas~yr$^{-1}$, in good agreement with the proper motion found by \citet{vasiliev_gaia_2021} from the Gaia EDR3 data directly. For the fast-moving SGB star we find a mean proper motion of of ($\mu_\alpha* , \mu_\delta$) = ($-3.37 \pm 0.45 , -5.87 \pm 0.45$) mas~yr$^{-1}$, well within the range of proper motions that we obtain for the other member stars and fully compatible with a cluster membership of this star. 
We therefore consider it likely that the SGB star is a cluster member.
Unfortunately we are not able to determine a proper motion for the main-sequence star.}

\paragraph*{Binary star} 
The orbital motion of stars in a binary system could be the source for the high radial velocity. Since we only have two radial velocities ($104.9 \pm 6.4$ and $99.1 \pm 5.4$~\kms{}) taken about 10 months apart for the subgiant star and only one for the main-sequence star, we cannot entirely exclude variations in the radial velocity. However, we note that our measurements are consistent with a constant radial velocity. 
Plausible binary system configurations that would produce a velocity amplitude of about 90~\kms{} containing a star with a mass of $0.75~\msol{}$ are a short-period system with a low-mass star, a system with a compact host (white dwarf, neutron star, stellar-mass black hole) with a period of several days, or a system with an IMBH as host with a period of several years.
\changed{To estimate the number of suitable binary systems,} we use a binary fraction of 5~per~cent estimated from our data, comparable to the low binary fraction of less then about 5~per~cent in this cluster \citep[see ][]{milone_acs_2012,ji_binary_2015}. \changed{This value is higher than the 2~per~cent of stars removed from our sample because of radial velocity variations (Section~\ref{sct:reliable}) since it is corrected for incompleteness due to a limited number of epochs and low velocity variations. Assuming} that 1~per~cent of all binary systems have a suitable configuration, we expect 0.5 of 1000 stars to be in such stellar binary systems.
If one of the stars is in a bound orbit around an IMBH with a mass of $4000~\msol{}$, its orbital distance must be less than $2 m_\text{IMBH} G / v^2$ which corresponds to 0.1~arcsec at the distance of \ngc{6093}.
This also excludes the possibility that both stars are bound to the same IMBH.

\paragraph*{Radial velocity outlier}
Since the radial velocities of both stars are about 40 and 50~\kms{} greater than the cluster escape velocity of 41.4~\kms{} \citep{baumgardt_catalogue_2018}, the stars are not gravitationally bound to the cluster. 

\paragraph*{Accelerated by interaction}
Interactions between binary systems with a single stellar-mass black hole can scatter one of the stars in the binary system and accelerate it to high velocities \citep[e.g. ][]{lutzgendorf_high-velocity_2012}. 
We expect the probability of this process to be proportional to the number of stellar-mass black holes in the cluster, which is low according to the best-fitting N-body models.
Similarly, an IMBH, if present, should be able to scatter stars in the same way as a stellar-mass black hole.

In conclusion, the mechanism leading to the high radial velocities is not obvious. One of the most likely explanations, the Kepler motion in a binary system, could easily be confirmed or rejected with a few additional measurements.

\section{Discussion}
\label{sect:discussion}

Our MCMC sampling reveals the existence of two possible solutions for the centre offsets of the Jeans model of \ngc{6093}: 
the northern solution has the centre close to the photometric centres from the literature and it does not need an IMBH to describe the observed kinematics, and a southern solution with a centre at a distance of about 2.4~arcsec from the northern one that needs an IMBH with a mass of $4600^{+1700}_{-1400}~\msol{}.$

The dynamical centre of the southern solution is at a distance of about 2 to 3~arcsec from the photometric centres in the literature.
A distinct dynamical centre would indicate a perturbed system, possibly caused by an IMBH.
We note that a distinct dynamical centre would be inconsistent with the assumption that the cluster can be described with a Jeans model where photometric and dynamical centre are the same by construction.
We try to minimize this discrepancy by choosing an MGE profile constructed for the current centre in each iteration.
Physically, the position of the IMBH is not expected to be identical with the photometric or dynamical centre.
$N$-body models predict that the IMBH will wander inside a sphere with a characteristic radius, the wandering radius $r_w$, around the centre. 
\citet{de_vita_wandering_2018} analysed $N$-body models and found a scaling relation for this radius which depends on a number of cluster parameters \citep[][Eq. 17]{de_vita_wandering_2018}.
Using values for the core density and radius from \citet{baumgardt_catalogue_2018} and assuming a mean stellar mass in the core of $0.65~\msol{}$, we find $r_w \approx 0\farcs{}4$ for an IMBH mass of about 3000 to $4600~\msol{}$. 
This radius is less than the distance between photometric and dynamical centre of the southern solution, indicating that the wandering motion can not be the explanation for the observed offset.
The sphere of influence of an IMBH with a mass of 4600~$\msol{}$ has a radius of 0.09~pc \citep{peebles_star_1972}, corresponding to 1.8~arcsec at the distance of \ngc{6093}.

\citet{lutzgendorf_limits_2013} found a $1\sigma$ upper limit of $800~\msol{}$ for an IMBH in \ngc{6093} using data from the integral-field spectrograph FLAMES/ARGUS. 
This value is well below all upper limits we derived.
\citet{lutzgendorf_limits_2013} calculated their own position of the cluster centre (see Fig.~\ref{fig:center}) and did not obtain the velocity dispersion by analysing spectra of individual stars, instead they combined unresolved spectra in radial bins. 
Their velocity dispersion profile is systematically lower in the cluster centre than ours: while our profile (Fig.~\ref{fig:dispersion}) is above 10~\kms{} for all radii below 10~arcsec (even after accounting for the uncertainty in the determination of the correct centre), their profile stays below 10~\kms{}. As discussed by \citet{bianchini_understanding_2015}, there are possible systematic errors in both methods, integrated-light spectroscopy and single-star kinematics. However, a common criticism regarding the latter is that the dispersion would be biased towards low values because of contamination from unresolved stars. The fact that our dispersion measurements are above those by \citet{lutzgendorf_limits_2013} suggests that our method is not affected by this. 
Our IMBH mass upper limit of $3000~\msol{}$ for the northern solution is below the IMBH mass estimate of $(3.63 \pm 0.95) \times 10^3\,\msol{}$ predicted from Monte-Carlo models \citep{arca_sedda_mocca-survey_2019}, while the median IMBH mass of the southern solution agrees with their result.
Estimates for the mass of an IMBH in \ngc{6093} based on the $M_\bullet$-$\sigma$ correlation and similar correlations range from 1000 to 2610~$\msol{}$ \citep[][Table 6]{safonova_extrapolating_2010}.


We assumed that the cluster kinematics can be described by an isotropic model (see Section~\ref{sect:model}). As \citet{zocchi_radial_2017} point out, radially anisotropic models of \ngc{5139} show an increase in the central velocity dispersion similar to that due to the influence of an IMBH.
If our assumption about isotropy is not satisfied in \ngc{6093}, our isotropic model would overestimate the IMBH mass. 
Since the cluster has a ratio of age to relaxation time greater than ten, the $N$-body models used in \citet{lutzgendorf_kinematic_2011} imply that the cluster centre is isotropic.

\changed{The comparison to $N$-body models performed in Section~\ref{sect:nbody} suggests that the observed steep surface brightness profile of the cluster is better explained by models without IMBH. The binned surface brightness profile presented in Figures~\ref{fig:dispersion}, \ref{fig:nbody_imbh} and \ref{fig:nbody_mbh} is derived from the photometry of \citet{dalessandro_peculiar_2018} which is based on HST observations. This profile is about 0.5 to 0.8~mag~arcsec$^{-2}$  brighter in the central bins compared to the profile of \citet{noyola_surface_2006} and about 1 to 1.3~mag~arcsec$^{-2}$ brighter than the profiles of \citet{trager_catalogue_1995} computed from ground-based photometry.
Figures~\ref{fig:nbody_imbh} and \ref{fig:nbody_mbh} indicate that \changedtwo{the other surface brightness profiles with lower central values is less well fitted by the no-IMBH model and better fitted by the models with an IMBH. }
An important difference between the two aforementioned surface brightness profiles and the one derived in this work is that the former are based on actual brightness measurements, whereas we adopted star counts. 
Profiles based on star counts are robust against shot noise effects caused by individual bright stars, yet require complete photometry even in the crowded cluster centres. The availability of \textit{ACS-HRC} data makes us confident that our results are not affected by incompleteness. Note that we only included stars brighter than \textit{F435W} $< 19.6$~mag, i.e. about the main-sequence turn-off, when determining the number density profiles.
}

Ultimately, our models cannot answer the question which of the two solutions for the cluster centre is to be preferred, and therefore whether an IMBH exists in M80. In light of the better agreement with the photometric determinations of the cluster centre, the northern solution seems the more likely one. 
However, our models do include a prior which favours solutions in agreement with the photometric centres. Hence, in case the true cluster centre coincides with the photometric estimates (and our northern solution), the high velocity dispersion around the centre of the southern solution is leading our Jeans models astray. As mentioned earlier, we verified that no individual high-velocity stars are responsible for the occurrence of the southern solution. Still, we cannot exclude that some of our model assumptions, such as a Gaussian line-of-sight velocity distribution or axisymmetry, are violated by the actual kinematics in the centre of M80.

\section{Conclusions}
\label{sect:conclusion}

We used spectra obtained with state-of-the-art adaptive optics of the central region in \ngc{6093} to analyse kinematic properties of the cluster.
We built on an axisymmetric Jeans model used previously for this cluster by \citet{kamann_peculiar_2020} and introduced additional parameters to describe a hypothetical intermediate-mass black hole and offsets between the photometric centre and the kinematic centre. 
The parameter samples of our Jeans model show a bimodal distribution in the centre offsets: a third of all samples are part of a northern solution with a centre close to the known photometric centres, the other two thirds are part of the southern solution with a centre at a distance of about 2.4~arcsec from the northern centre.
While most parameters do not show significant differences between the two solutions, the distribution of mass of the central IMBH is different.
Around the northern centre, we find a distribution with a peak below $1000~\msol{}$ that quickly decreases with increasing IMBH mass. The 90-per-cent upper limit is $3000~\msol{}.$
The IMBH mass distribution of the southern solution has a median and an $1\sigma$ uncertainty of $4600^{+1700}_{-1400}~\msol{}$, 90~per~cent of all samples are below $6800~\msol{}.$
$N$-body models support the existence of an IMBH in this cluster with a mass of up to $6000~\msol{}$ \changed{although models without an IMBH provide a better fit to the observed surface brightness profile. They further indicate that} the cluster has lost nearly all stellar-mass black holes.
The overall radial profiles of the mass-to-light-ratio and the rotation velocity agree with the previous analysis of this cluster \citep{kamann_peculiar_2020}.
While the part outside the cluster core clearly rotates, it is not clear whether the core of the cluster is rotating. Our analysis is consistent with no rotation in the centre but the uncertainty in the position of the cluster centre prohibits any definitive conclusion.
We discussed the detection of two central stars with radial velocities clearly above the escape velocity of the cluster.
Their high velocities could be explained if one or both stars belong to the Milky Way population instead to \ngc{6093}, if they are caused by binary motion, or if one or both stars were accelerated by interactions between a binary system and a stellar-mass or intermediate-mass black hole. Of these possibilities, the periodic change of radial velocity due to binary motion can be confirmed or ruled out by future observations.

Proper-motion data of stars in the cluster centre could complement the radial velocities used in this study.
Deep radio and X-ray observations of the cluster centre could lead to further insights about a possible IMBH in \ngc{6093}, similar to other clusters \citep{tremou_maveric_2018}. Finally, detailed studies of the stellar mass function or the binary fraction within the core radius of M80 could be used to understand the amount of mass segregation present near the centre of M80. As argued by some authors, e.g. \citet{gill_intermediate-mass_2008}, an IMBH would strongly reduce the amount of mass segregation expected near the centre. In this respect, the finding of \citet{dalessandro_peculiar_2018} of differences in the concentrations of the three chemically distinct populations discovered in M80 might be considered as evidence against an IMBH.

\section*{Data availability}
The data underlying this article are available in the Göttingen Research Online repository, at \url{https://doi.org/10.25625/VCNHOR}.
The raw exposures used in this article are available under their Program ID (Table~\ref{tab:observations}) in the ESO Archive, at \url{https://archive.eso.org}.

\section*{Acknowledgements}
\changed{We thank the referee Glenn van de Ven for his constructive suggestions and comments that helped to improve the paper.}
FG, SD, BG, and TOH acknowledge funding from the Deutsche Forschungsgemeinschaft (grant DR~281/35-1 and KA~4537/2-1) and from the German  Ministry for Education and Science (BMBF-Verbundforschung) through grants 05A14MGA, 05A17MGA, 05A14BAC, 05A17BAA and 05A20MGA. SK gratefully acknowledges funding from UKRI in the form of a Future Leaders Fellowship (grant no. MR/T022868/1).
The authors acknowledge the French National Research Agency (ANR) for supporing this work through the ANR APPLY (grant ANR-19-CE31-0011) coordinated by B. Neichel.
The work was supported by the North-German Supercomputing Alliance (HLRN).




\bibliographystyle{mnras}
\bibliography{ngc6093_paper}



\appendix

\section{Additional figures and tables}

\begin{figure}
	\includegraphics[width=\columnwidth]{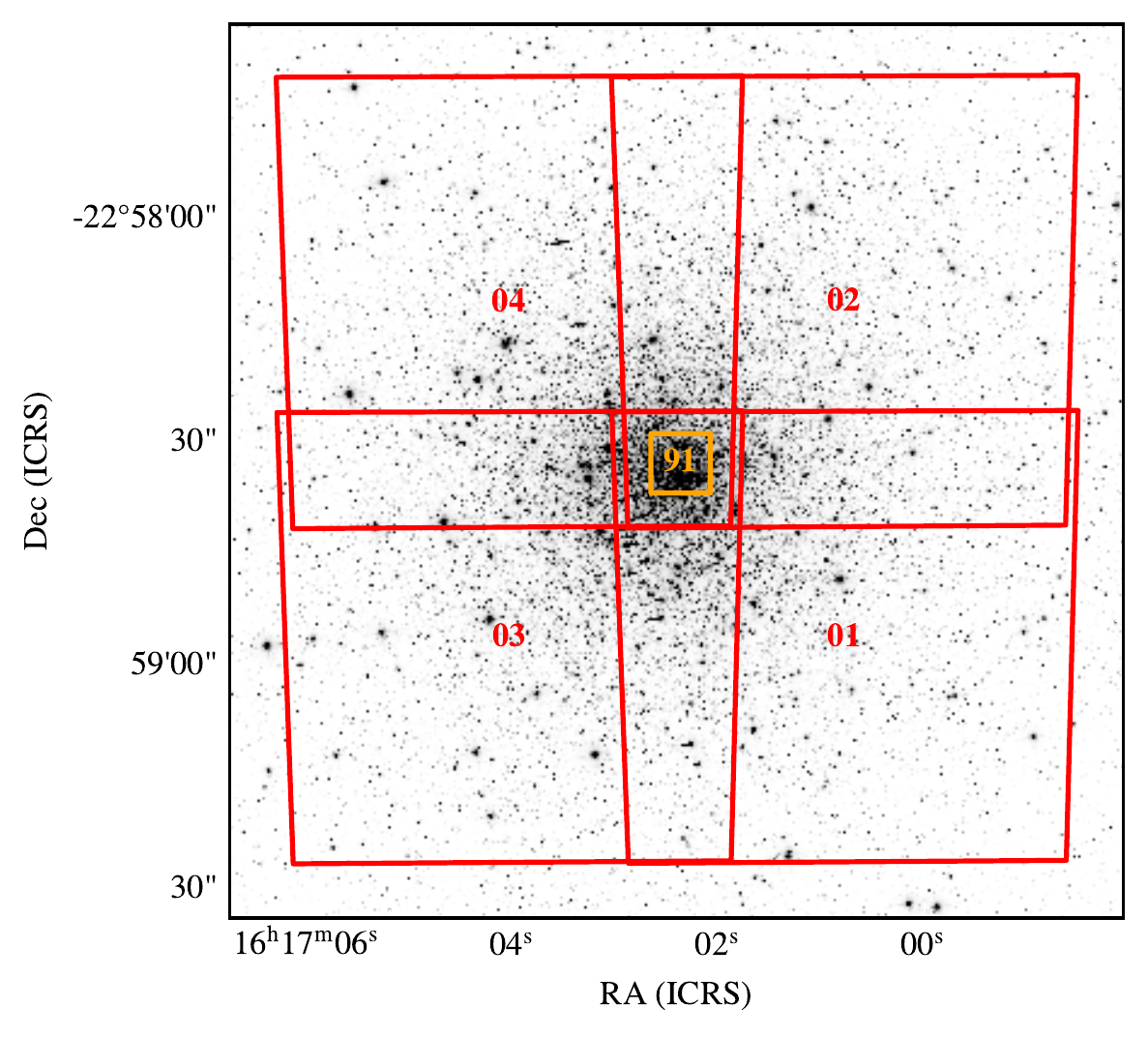}
    \caption{The four MUSE WFM pointings (01--04) and the central NFM pointing (91) on top of the \hst{} \textit{ACS} image of \ngc{6093} \citep{sarajedini_acs_2007,anderson_acs_2008}.}
    \label{fig:pointings}
\end{figure}

\begin{figure*}
	\includegraphics[width=1.45\columnwidth]{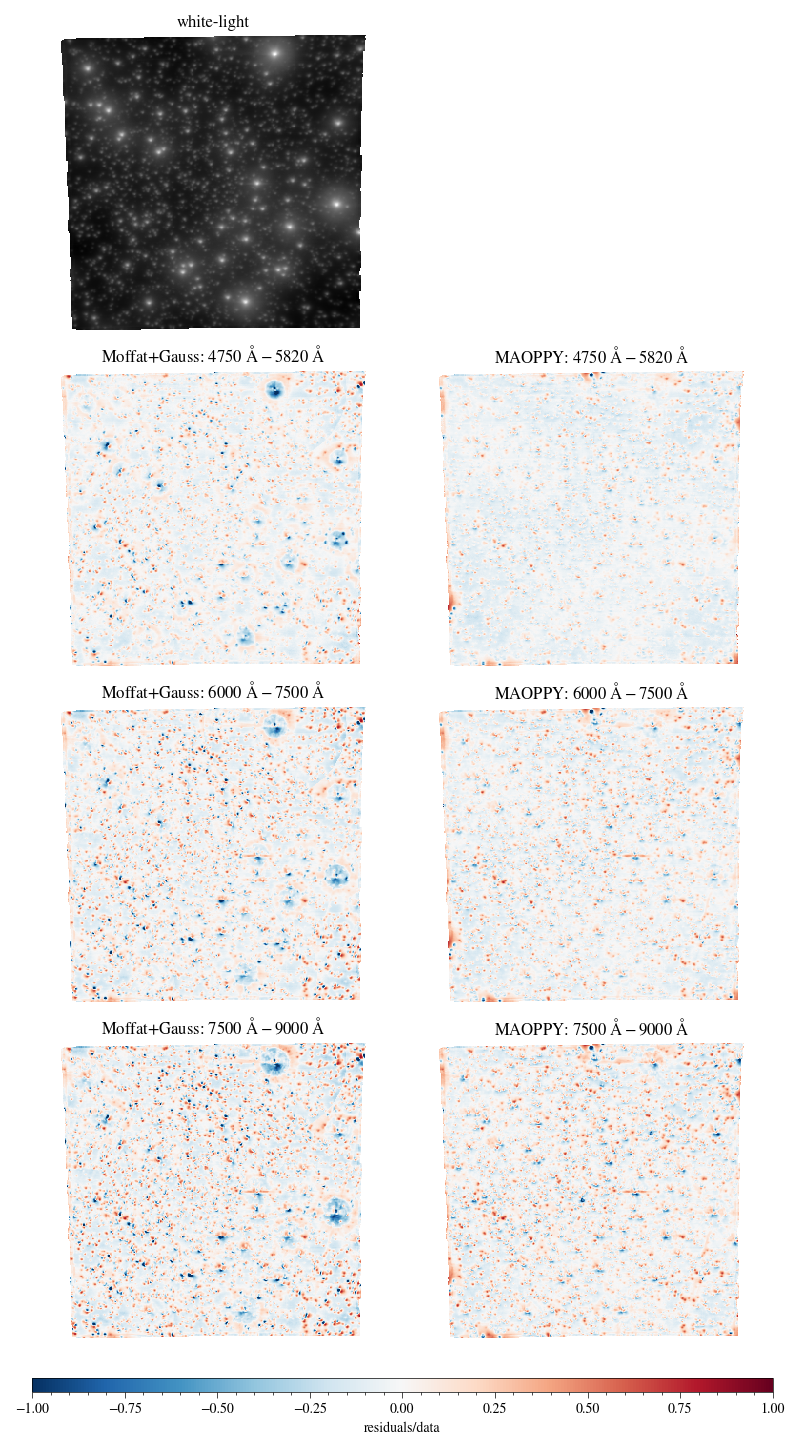}
    \caption{White-light image of a NFM observation and the residuals relative to the original data after subtracting the PSF model (\changed{left column}: a combination of a Moffat curve with a Gaussian, \changed{right column}: \textsc{Maoppy}) and background in three different wavelength ranges.}
    \label{fig:psf_residuals}
\end{figure*}

\begin{figure*}
	\includegraphics[width=2\columnwidth]{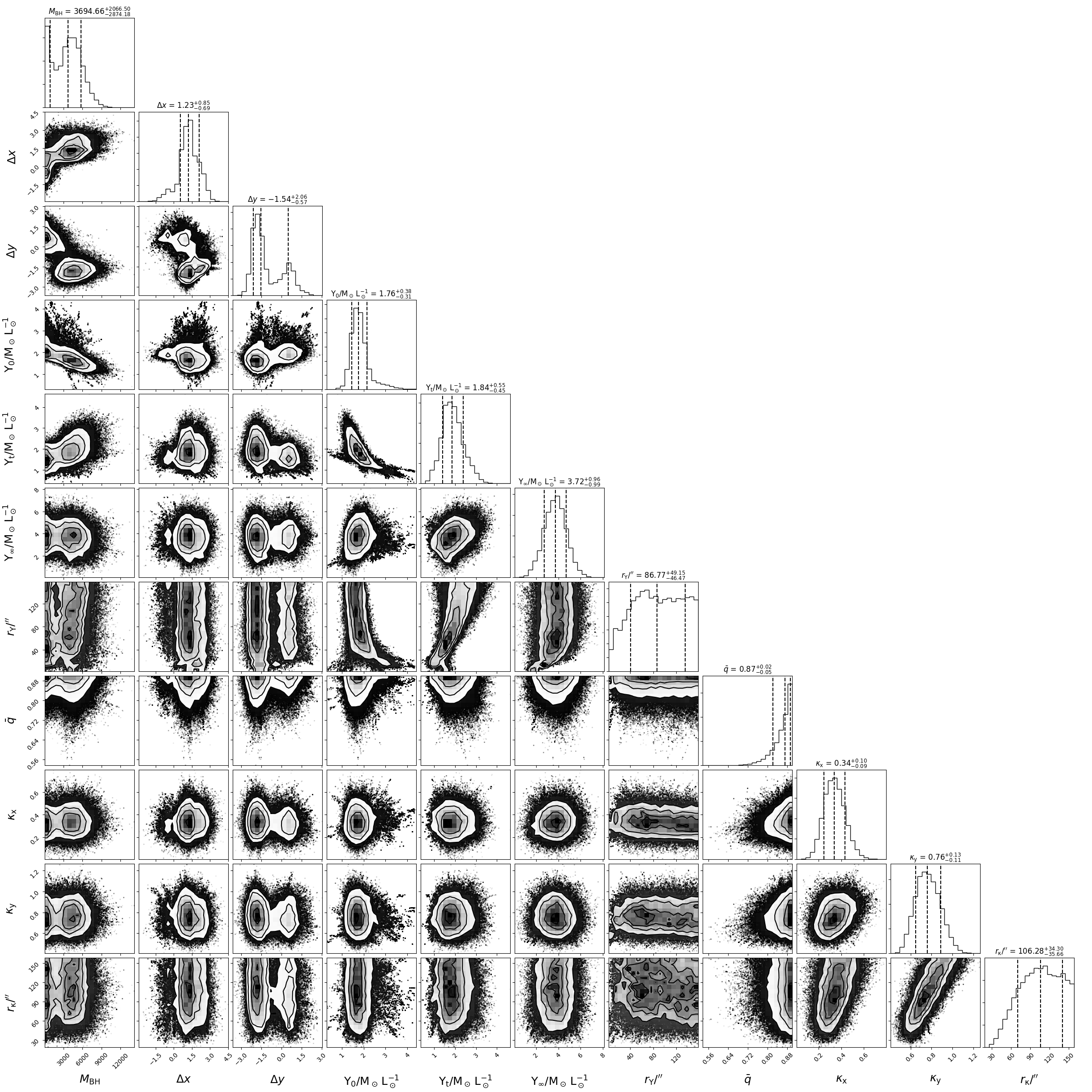}
    \caption{Corner plot of our final MCMC chain. See Table~\ref{tab:priors} for a description of the parameters.}
    \label{fig:corner}
\end{figure*}

\begin{figure}
	\includegraphics[width=\columnwidth]{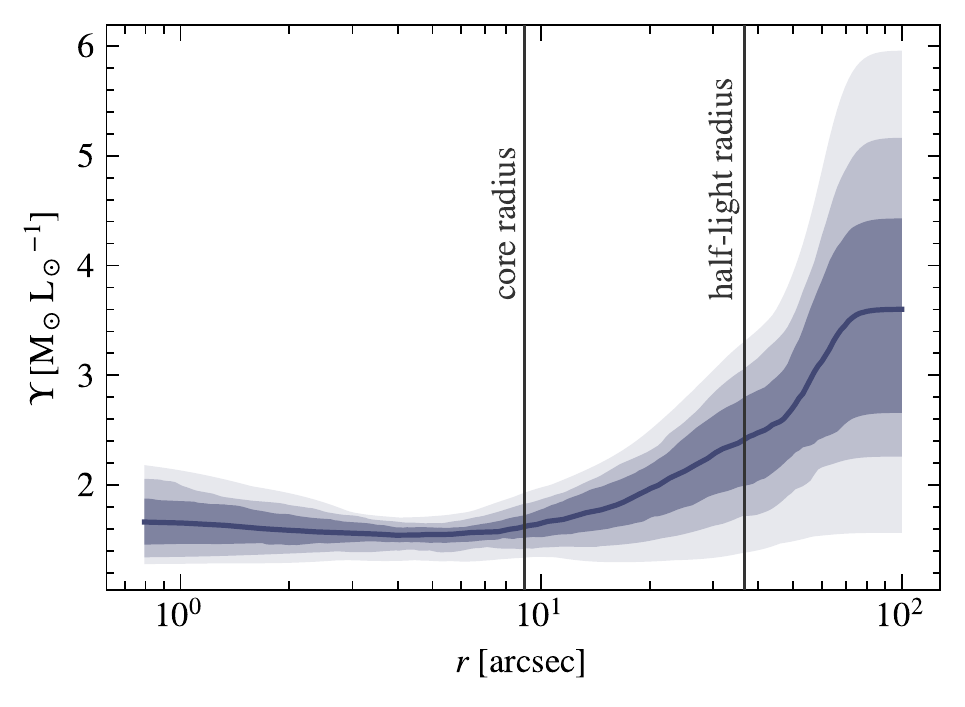}
    \caption{Radial profiles of the mass-to-light ratio $\Upsilon$, the blue lines corresponds to the median, the shaded areas are the $1\sigma$, $2\sigma$, and $3\sigma$ limits.}
    \label{fig:mlr}
\end{figure}

\begin{figure}
	\includegraphics[width=\columnwidth]{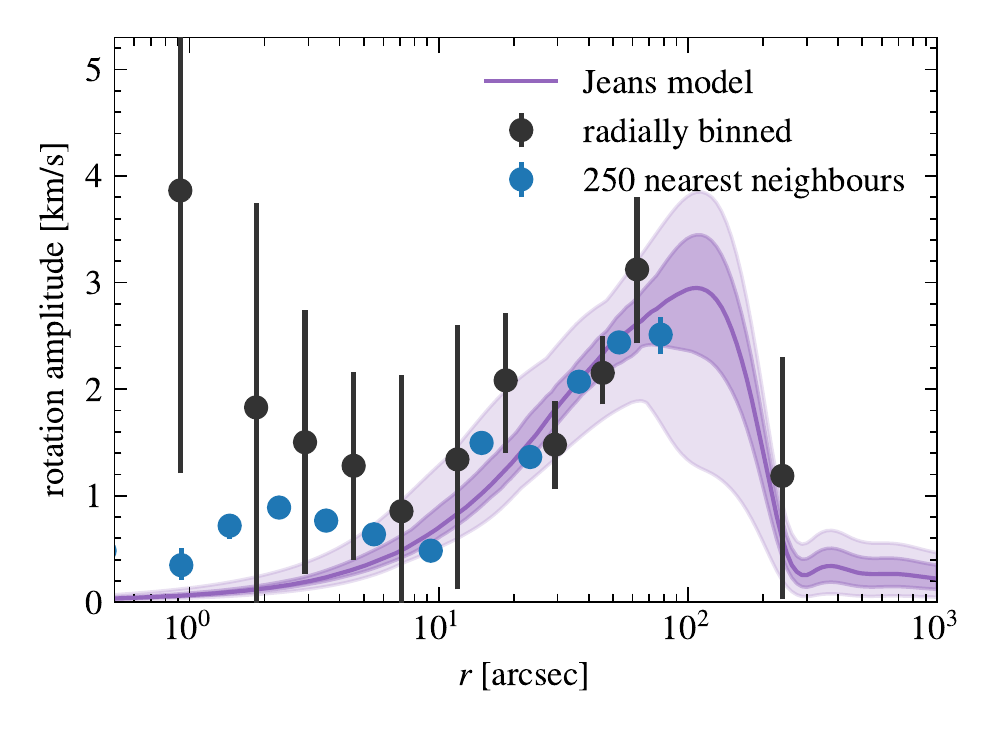}
    \caption{Radial profile of the rotation amplitude \changedtwo{derived from the Jeans model, from individual radial velocities in radial bins, and from the mean radial velocities of the 250 nearest neighbours of each star. }}
    \label{fig:rotation}
\end{figure}

\begin{table*}
	\centering
	\caption{MUSE observations of \ngc6093.}
	\label{tab:observations}
        \begin{tabular}{cclcr}
        \hline
        Pointing &  Obs. date &   Inst. mode &             Exp. time  & Prog. ID \\
        \hline
        01 &  2015-05-11 07:56:56 &     WFM &  $3 \times 200$~s & 095.D-0629  \\
         &  2017-04-23 09:10:55 &     WFM &  $3 \times 200$~s & 099.D-0019  \\
         &  2019-05-04 06:44:49 &  WFM-AO &  $3 \times 200$~s & 0103.D-0204 \\
        \hline
        02 &  2015-05-11 08:12:39 &     WFM &  $3 \times 200$~s & 095.D-0629  \\ 
         &  2017-02-01 09:11:41 &     WFM &  $3 \times 200$~s & 098.D-0148  \\
         &  2017-04-23 09:29:43 &     WFM &  $3 \times 200$~s & 099.D-0019  \\
         &  2019-05-04 07:01:07 &  WFM-AO &  $3 \times 200$~s & 0103.D-0204 \\
        \hline
        03 &  2015-05-11 08:42:29 &     WFM &  $3 \times 200$~s & 095.D-0629  \\ 
         &  2015-05-11 08:58:28 &     WFM &  $3 \times 200$~s & 095.D-0629  \\ 
         &  2017-04-26 04:22:07 &     WFM &  $3 \times 200$~s & 099.D-0019  \\
         &  2019-05-04 07:25:53 &  WFM-AO &  $3 \times 200$~s & 0103.D-0204 \\
        \hline
        04 &  2015-05-11 09:14:19 &     WFM &  $3 \times 200$~s & 095.D-0629  \\
         &  2017-04-26 04:37:08 &     WFM &  $2 \times 200$~s & 099.D-0019  \\
         &  2019-05-04 07:42:12 &  WFM-AO &  $3 \times 200$~s & 0103.D-0204 \\
        \hline
         91 &  2019-05-04 09:40:10 &  NFM-AO &  $4 \times 600$~s & 0103.D-0204 \\
         &  2020-02-24 08:45:05 &  NFM-AO &  $4 \times 600$~s & 0104.D-0257 \\
        \hline
        \end{tabular}
\end{table*}

\begin{table*}
	\centering
	\caption{Priors used in the Jeans model MCMC.}
	\label{tab:priors}
        \begin{tabular}{cccl}
        \hline
        Parameter & Prior & Unit & Description \\
        \hline
        $m_\text{IMBH}$ & $\text{uniform}(0, 15)$ & $10^3 \msol$ & mass of central dark component \\
        $\Delta x ,\Delta y$ & $\text{normal}(\mu=0, \sigma=1)$ & arcsec & offset of GC centre position  \\
        $\Upsilon_0$ & $\text{uniform}(0.1, \infty)$ & $\msol \lsol{}^{-1}$ & central mass-to-light ratio \citep{kamann_peculiar_2020} \\
        $\Upsilon_\text{t}$ & $\text{uniform}(0.1, \infty)$ & $\msol \lsol{}^{-1}$ & mass-to-light ratio at $r = r_\Upsilon$ \\
        $\Upsilon_\infty $ & $\text{normal}(\mu=3.5, \sigma=1)$  & $\msol \lsol{}^{-1}$ & mass-to-light ratio at large radii \\
        $r_\Upsilon$ & uniform(min($\sigma_i$), max($\sigma_i)$)$^1$ & arcsec & radius at which $\Upsilon(r=r_\Upsilon) = \Upsilon_\text{t}$\\
        $\bar{q}$ & $\text{uniform}(0.2, \text{median}(q_i))$ & dim. less & parametrization of inclination$^1$ \citep{watkins_discrete_2013} \\
        $\kappa_x , \kappa_y$ &  $\text{normal}(\mu=0, \sigma=5)$  & km s$^{-1}$ & components of rotation vector \\
        $r_\kappa$ & uniform(min($\sigma_i$), max($\sigma_i$)) & arcsec & scaling length for rotation profile \\
        
        \hline 
        \multicolumn{4}{l}{$^1$ $\sigma_i$ and $q_i$ are the width and the axial ratio of the $i$-th component of the MGE.}\\
        \end{tabular}
\end{table*}

\begin{table}
	\centering
	\caption{Fit statistics for $N$-body models around the northern centre. `RV' refers to fits of the binned radial velocities, `SB' to fits of the surface brightness profile, and `total' to simultaneous fits to both.}
	\label{tab:fits_north}
        \begin{tabular}{rrrrr}
        \hline
        $m_\text{IMBH}$ [$\msol{}$] & SBH retention frac. [\%] & $\chi{}_r^2$ RV & $\chi_r^2$ SB & $\chi_r^2$ total  \\
        \hline
        0 & 10 & 1.23 & 0.78 & 0.93 \\
        0 & 30 & 2.43 & 1.26 & 1.65 \\
        0 & 50 & 1.79 & 1.57 & 1.93 \\
        0 & 100 & 1.16 & 1.68 & 2.38 \\
        1500 & 10 & 1.94 & 1.56 & 1.66 \\
        3000 & 10 & 1.68 & 1.42 & 1.49 \\
        6000 & 10 & 1.14 & 1.07 & 1.09 \\
        15000 & 10 & 2.10 & 1.33 & 1.54 \\
        \hline
        \end{tabular}
\end{table}

\begin{table}
	\centering
	\caption{Similar to Table~\ref{tab:fits_south} but for $N$-body models around the southern centre.}
	\label{tab:fits_south}
        \begin{tabular}{rrrrr}
        \hline
        $m_\text{IMBH}$ [$\msol{}$] & SBH retention frac. [\%] & $\chi{}_r^2$ RV & $\chi_r^2$ SB & $\chi_r^2$ total  \\
        \hline
        0 & 10 & 2.25 & 0.94 & 1.24 \\
        0 & 30 & 3.35 & 1.15 & 1.82 \\
        0 & 50 & 2.34 & 1.47 & 2.04 \\
        0 & 100 & 1.77 & 1.58 & 2.51 \\
        1500 & 10 & 2.00 & 1.33 & 1.51 \\
        3000 & 10 & 2.06 & 0.97 & 1.21 \\
        6000 & 10 & 1.54 & 1.02 & 1.16 \\
        15000 & 10 & 1.87 & 1.29 & 1.44 \\
        \hline
        \end{tabular}
\end{table}


\bsp	
\label{lastpage}
\end{document}